\documentclass[twocolumn]{aastex62}

\usepackage[single=true]{acro}

\DeclareAcronym{dsa}{
  short = DSA ,
  long = diffusive shock acceleration,
  class = astro ,
  first-style = default
}
\DeclareAcronym{snr}{
  short = SNR ,
  long  = supernova remnant ,
  class = astro ,
  first-style = default
}
\DeclareAcronym{nw}{
  short = NW ,
  long  = northwest ,
  class = astro ,
  first-style = default
}
\DeclareAcronym{hxc}{
  short = HXC ,
  long  = hard X-ray component ,
  class = astro ,
  first-style = default
}
\DeclareAcronym{gcr}{
  short = GCR ,
  long  = Galactic cosmic ray ,
  class = astro ,
  first-style = default
}
\DeclareAcronym{psf}{
  short = PSF ,
  long  = point spread function ,
  class = astro ,
  first-style = default
}
\DeclareAcronym{fov}{
  short = FOV ,
  long  = field of view ,
  class = astro ,
  first-style = default
}
\DeclareAcronym{cta}{
  short = CTA ,
  long  = Cherenkov Telescope Array ,
  class = astro ,
  first-style = default
}
\DeclareAcronym{mhd}{
  short = MHD ,
  long  = magnetohydrodynamical ,
  class = astro ,
  first-style = default
}
\DeclareAcronym{ism}{
  short = ISM ,
  long  = interstellar medium ,
  class = astro ,
  first-style = default
}
\DeclareAcronym{ic}{
  short = IC ,
  long  = inverse Compton ,
  class = astro ,
  first-style = default
}

\received{MM DD, YYYY}    \revised{MM DD, YYYY}    \accepted{MM DD, YYYY}

\shorttitle{Hotspots in the NW shell of the SNR RX J1713.7-3946}
\shortauthors{Higurashi et al.}

\newcommand{\rxj}{RX J1713.7$\--$3946}
\newcommand{\rxjG}{G347.3$-$0.5}

\newcommand{\hs}{hotspot}
\newcommand{\hsS}{hotspots}
\newcommand{\HS}{Hotspot}

\newcommand{\chandra}{{\it Chandra}}
\newcommand{\rosat}{{\it ROSAT}}
\newcommand{\nustar}{{\it NuSTAR}}
\newcommand{\xmm}{{\it XMM-Newton}}
\newcommand{\suzaku}{{\it Suzaku}}
\newcommand{\hess}{H.E.S.S.}

\newcommand{\lat}{{\it Fermi}-LAT}

\newcommand{\figref}[1]{Figure~\ref{#1}} 
\newcommand{\tabref}[1]{Table~\ref{#1}} 
\newcommand{\eqref}[1]{Eq.~(\ref{#1})} 
\newcommand{\secref}[1]{Section~\ref{#1}} 

\begin{document}

\title{X-RAY HOTSPOTS IN THE NORTHWEST SHELL OF THE~SUPERNOVA~REMNANT RX J1713.7$\--$3946}

\email{r.higurashi@rikkyo.ac.jp, naomi.tsuji@riken.jp, y.uchiyama@rikkyo.ac.jp}

\author{Ryota Higurashi}
\affil{Department of Physics, Rikkyo University, 3-34-1 Nishi Ikebukuro, Toshima-ku, Tokyo 171-8501, Japan}

\author{Naomi Tsuji}
\affiliation{Interdisciplinary Theoretical \& Mathematical Science Program (iTHEMS), RIKEN, 2-1 Hirosawa, Saitama 351-0198, Japan}

\author{Yasunobu Uchiyama}
\affiliation{Department of Physics, Rikkyo University, 3-34-1 Nishi Ikebukuro, Toshima-ku, Tokyo 171-8501, Japan}



\begin{abstract}
The supernova remnant (SNR) \rxj\ is one of the best-studied accelerators of cosmic rays because of its strong nonthermal X-ray and gamma-ray radiation.
We have analyzed accumulated \chandra\ observations with a total exposure time of $\sim$266 ks in the northwest rim of \rxj.
We detect a substantially large number of point-like sources, referred to as ``\hsS'', which are likely associated with the remnant.
The spectra of the \hsS\ are well described by an absorbed power-law model.
The spectral properties ($10^{21}\ \mathrm{cm^{-2}}\lesssim N_H \lesssim 10^{23}\ \mathrm{cm^{-2}}$ and $ 0.5\lesssim \Gamma \lesssim 6$) are different from diffuse X-ray emission in \rxj, and the harder \hs\ tends to have the larger $N_H$.
We also confirm yearly and monthly variabilities of flux for some \hsS.
We propose that \rxj\ is embedded in a complex surroundings where some dense molecular clumps and cores exist inside a wind-blown cavity, and
that the \hs\ traces synchrotron emission caused by an interaction of shock waves of the SNR and dense molecular cores
with a number density of $10^{5}\--10^{7}~ \mathrm{cm}^{-3}$.
The X-ray radiation of the \hs\ might be emitted from both primary electrons accelerated at the shocks and secondary electrons produced by the interaction of accelerated protons with the cores.

\end{abstract}



\section{Introduction} \label{sec:Intro}

Supernova remnants are considered as the primary accelerators of the \acp{gcr} with energies smaller than the ``knee'' (a few PeV).
The well-studied theory of \ac{dsa} \citep{Axford1977a,Krymskii1977a,Bell1978b,Blandford1978a} is widely accepted for acceleration at shock waves of the \ac{snr}.
\citet{Koyama1995a} first discovered synchrotron X-ray emission in the energy range of $0.4\--8.0\ \mathrm{keV}$ radiated from electrons accelerated up to multi-TeV energies in the shell of SN~1006.
The subsequent detection of synchrotron X-rays from other \ac{snr}s, including \rxj\ \citep{Koyama1997a}, makes \ac{snr}s among the best candidates for studying particle acceleration.
Gamma-ray emission from \ac{snr}s that include \rxj\ is also detected in the range of GeV$\--$TeV energy (e.g., \citet{Funk2015a} and references therein), and this supports particle acceleration in \ac{snr}s.
Accelerated protons produce gamma rays via decay of neutral pions produced by collisions with ambient protons or nuclei (``hadronic scenario'').
Accelerated electrons might also generate gamma rays by \ac{ic} scattering of low-energy photons (``leptonic scenario'').
Gamma-ray radiation from accelerated protons was revealed by the detection of so-called ``pion bumps'' in middle-aged \ac{snr}s interacting with molecular clouds \citep{Ackermann2013a,Jogler2016a},
whereas no obvious evidence has yet been found that protons are accelerated to the knee in \ac{snr}s.

The shell-type \ac{snr} \rxj\ (also known as \rxjG) was first discovered by \rosat\ All-Sky Survey \citep{Pfeffermann1996a}.
The distance was estimated to be $\sim 1\ \mathrm{kpc}$ by tracing the CO lines from the associated molecular clouds  \citep{Fukui2003a}.
Its association with one of the historical SNRs, namely SN393, has been discussed \citep{Wang1997a,Fesen2012a}.
Recent measurements of the forward shock velocity suggested the SN--SNR association, supporting an \ac{snr} age of $\sim 1600\ \mathrm{yr}$ \citep{Tsuji2016a,Acero2017a}.
\rxj\ is one of the best-studied objects for investigating particle acceleration because of its strong non-thermal X-ray and gamma-ray emissions.

The X-ray emission of \rxj\ is dominated by non-thermal radiation, i.e., synchrotron radiation that is emitted by very-high-energy electrons accelerated in the \ac{snr} shock via \ac{dsa} \citep{Koyama1997a,Slane1999a,Uchiyama2003a,Cassam-Chenai2004a,Hiraga2005a,Takahashi2008a,Tanaka2008a}.
The acceleration efficiency was found to be close to the maximum rate (Bohm limit) 
based on the X-ray observations \citep{Uchiyama2007a,Tanaka2008a,Tsuji2019a}.
\cite{Katsuda2015a} detected thermal X-ray components in the inner region.
They inferred that the thermal line emission originates from the ejecta heated by the reverse shock, and the mass of the progenitor is estimated to be $\lesssim 20\ M_{\sun}$ by the composition of its ejecta.
In the shell region, complex filamentary structures have been revealed by \chandra\ with its superb angular resolution (0.5\arcsec) \citep{Uchiyama2003a,Lazendic2004a}.
\citet{Okuno2018a} lately presented the spatially resolved spectroscopy of the \ac{snr} with \chandra.
They showed that the filamentary structures in the southeast (SE) have harder spectra ($\Gamma\sim 2.0$) than the surrounding regions, whereas those in the southwest (SW) have relatively soft spectra ($\Gamma\sim 2.7$) because of  the deceleration of shock waves which are interacting with the clumpy \ac{ism} produced by the stellar wind of the massive progenitor star.
Recently, \nustar\ observations of the \ac{nw} rim obtained a spatially resolved X-ray image in the 10$\--$20 keV energy band \citep{Tsuji2019a}.
The hard X-ray morphology is roughly in agreement with the soft-band images of previous works.

\rxj\ is known as a strong TeV gamma-ray emitter \citep{Aharonian2006a,Aharonian2007a}, but the radiation mechanism (``hadronic'' and/or ``leptonic'') has been in a matter of debate \citep[e.g., ][]{Uchiyama2003a,Aharonian2006a,Tanaka2008a,Ellison2010a,Zirakashvili2010a,Abdo2011a,Fukui2012a,Inoue2012a,Sano2013a,Gabici2014a,Celli2019a,H.E.S.S.Collaboration2018a}.
The spatial coincidence between the non-thermal X-ray and the TeV gamma ray may imply that the parent particles responsible for these two radiations are identical, i.e., electrons.
The leptonic radiation also seems to be favored by a hard spectrum in the GeV energy band observed with \lat\ \citep{Abdo2011a}.
The observed photon index ($\Gamma = 1.5\pm0.1$) in the bandpass of \lat\ is reconciled with the \ac{ic} gamma ray radiated from the electron population with a spectral index of 2, which is expected from standard acceleration via \ac{dsa}.
Note that a simple modeling with unique population of electrons fails to reproduce the multi-wavelength spectrum \citep{H.E.S.S.Collaboration2018a}.
It has lately been shown that the observed GeV$\--$TeV gamma-ray spectrum can be reproduced by the hadronic scenario as well \citep{Inoue2012a,Gabici2014a,Celli2019a}.
Because \rxj\ is located in a complex region in the Galactic Plane, there exist target materials to produce $\pi^0$-decay gamma rays through the pp interaction (see below for details).
Although deeper \hess\ observations with total live time of 164 h have been conducted, the mechanism of the gamma-ray radiation remains ambiguous \citep{H.E.S.S.Collaboration2018a}.
There are several ways to distinguish the gamma-ray origin, such as by detection of synchrotron emission from a secondary electron produced by the decay of a charged pion in the pp interaction \citep{Huang2018a} or by accurate observation around 100 TeV with \ac{cta} because the Klein--Nishina effect would suppress the leptonic emission component \citep{Acero2017b}.
The confirmation of the hadronic component would also come from observations of neutrinos, with large volume instruments as KM3NeT\citep{Ambrogi2018a}.

\rxj\ is located in the complex environment of ambient molecular clouds \citep{Fukui2003a,Fukui2012a}.
The progenitor star is expected to have exploded in a cavity wall produced by the stellar wind from the massive progenitor star.
In this case, dense materials, such as molecular clumps and molecular cores, could have survived against the wind.
Here, we refer to a ``clump" as a denser part inside a molecular cloud with a number density of $10^{3}\--10^{4}\ \mathrm{cm^{-3}}$ and a typical size of 0.1 pc,
and a ``core’’ as a much denser part inside the clump with a number density of $10^{5}\--10^{7}\ \mathrm{cm^{-3}}$ and a typical size of 0.01 pc.
The medium with a lower density is thought to be swept up before the SN explosion.
This results in the complex surrounding which has some molecular clumps and cores inside the wind-blown cavity,
as mentioned in \citet{Slane1999a,Inoue2012a,Katsuda2015a}.

Shock--cloud interactions in such complex circumstances have been studied with \ac{mhd} simulations.
\citet{Inoue2012a} revealed that the magnetic field is amplified up to $1\ \mathrm{mG}$ in a small region ($\sim 0.05\ \mathrm{pc}$), resulting from the SNR shock propagating within the clumpy environment.
Recently, \citet{Celli2019a} showed that magnetic field amplification occurs at a thin region (referred to as ``skin’’) with a size of $\sim$0.05 pc at the surface of the molecular clump.

Observations have been in agreement with the picture of the shock--cloud interaction in \rxj \citep{Fukui2003a,Fukui2012a,Moriguchi2005a,Sano2010a,Sano2013a}.
Using observations with \suzaku, \citet{Sano2015a} reported a parsec-scale correlation between the intensities of the non-thermal X-ray emission and the molecular clumps.
They also found an anti-correlation between the X-ray radiation and the molecular clumps on a sub-parsec scale.
This suggested that the synchrotron X-ray intensity is indeed strong in the surroundings of the clumps where the magnetic field is expected to be enhanced.
\citet{Okuno2018a} showed a similar anti-correlation between molecular clumps and the photon indices of the X-ray spectra with \chandra.
The presence of a magnetic field with $B\sim 1\ \mathrm{mG}$ was reported in small knot-like regions in the \ac{nw} shell of \rxj, inferred from the detection of year-scale variability caused by a balance between the synchrotron emission and the acceleration \citep{Uchiyama2007a}.

Here, we report the detection of a significantly larger population of point-like sources in the \ac{nw} rim of \rxj.
In \secref{sec:Obs}, we summarize the \chandra\ observations. 
Detailed analyses and results are presented in \secref{sec:AnalysisAndResults}.
We discuss the physical meaning of the newly found sources in \secref{sec:Discussions}.
The conclusions are given in \secref{sec:conclusion}.

\begin{figure}[t!]
\plotone{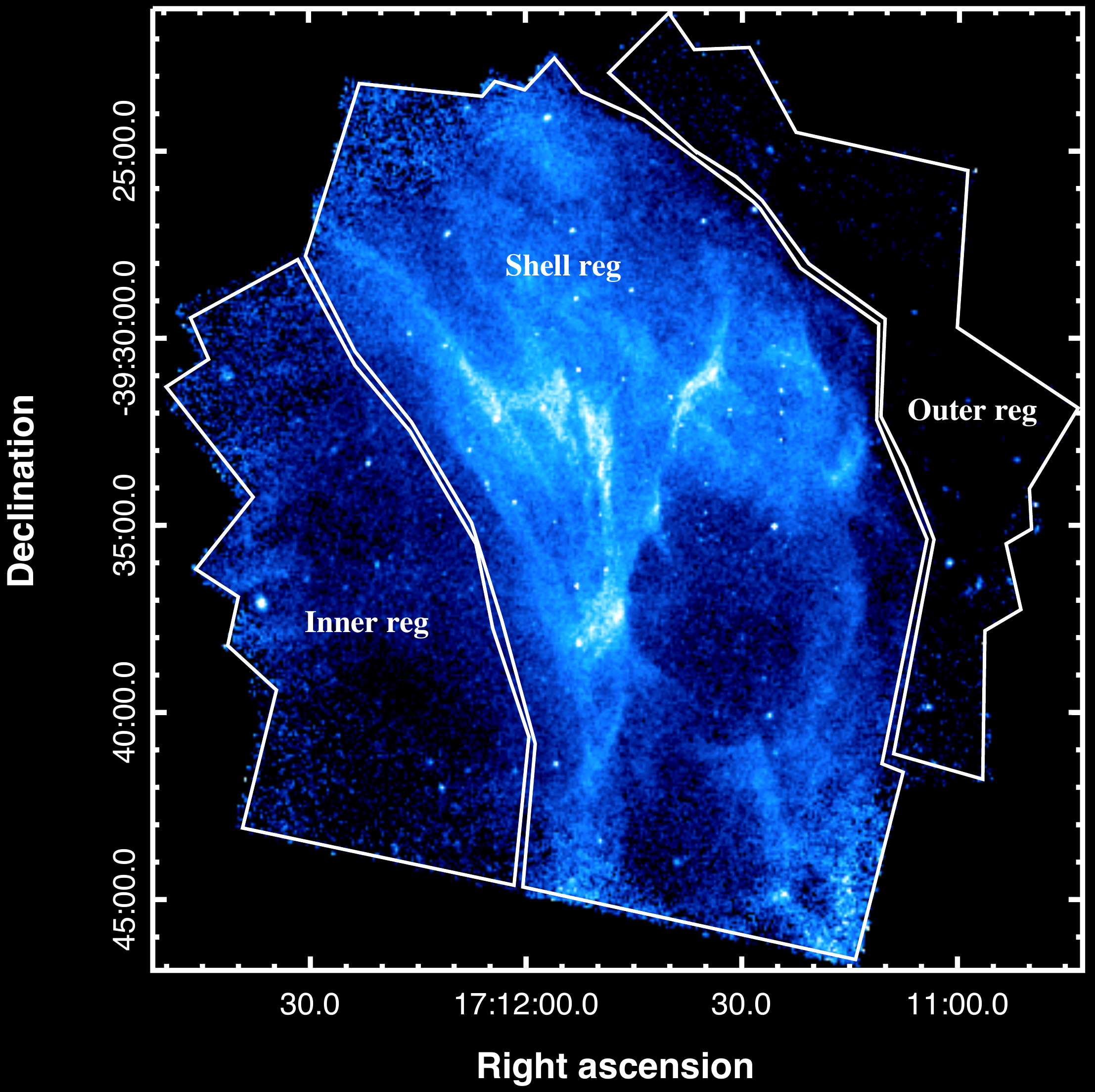}
\caption{
The combined flux image of \rxj\ \ac{nw} at $0.5\--7.0\ \mathrm{keV}$.
Each pixel has a width of $1\ \mathrm{arcsec}$. The three white regions indicate ``Inner reg'', ``Shell reg'', and ``Outer reg''.
}
\label{fig:1713_regs}
\end{figure}

\section{Observations and Data Reduction} \label{sec:Obs}

We performed observations of \rxj\ \ac{nw} with the \chandra\ Advanced CCD Imaging Spectrometer (ACIS)-I six times from 2005 to 2011 (\tabref{tab:chandra_dataset}).
We also made use of archival data taken in 2000.
The total exposure time reached $\sim 266\ \mathrm{ks}$, which allowed us to investigate relatively faint structures.
Moreover, the seven-time observations over a span of eleven years, from 2000 to 2011, enabled us to explore both year-scale and month-scale variability by utilizing the three observations taken in 2009.
Taking advantage of the great angular (sub-arcsecond) resolution of \chandra\ and the accumulated rich statistics,
we performed a detailed analysis of small and characteristic structures in the \ac{nw} rim of \rxj.
All the data were reprocessed using {\tt chandra\_repro} with CALDB version 4.7.6 in Chandra Interactive Analysis of Observations (CIAO) version 4.9, provided by the Chandra X-Ray Center (CXC).

\begin{deluxetable*}{ccccccc}[ht!]
\tablecaption{\chandra\ observations of \rxj\ \ac{nw}, CTB~37A and IC~4637 }
\tablehead{
 \colhead{ObsID} & \colhead{Target name} & \colhead{Start date} & \twocolhead{Pointing position} &  \colhead{Exposure} & \colhead{A.D.\tablenotemark{$\dagger$}}  \\
\colhead{}& \colhead{} & \colhead{(yyyy-mm-dd)} & \colhead{($\alpha\ [J2000],\ \delta\  [J2000]$)} & \colhead{($l,\ b$)\tablenotemark{$\ast$}}  & \colhead{(ks)} &\colhead{(arcmin)}
}
\startdata
\hline
736    & \rxj\ \ac{nw} & 2000-07-25 &  17:11:49.9, $-$39:36:14.7 & 347.2660, $-$0.1121 &  29.6    & -- \\
5560   & \rxj\ \ac{nw} & 2005-07-09 &  17:11:45.5, $-$39:33:40.0 & 347.2923, $-$0.0752 &  29.0  & --\\
6370   & \rxj\ \ac{nw} & 2006-05-03 &  17:11:46.3, $-$39:33:12.0 & 347.3001, $-$0.0728 &  29.8  & --\\
10090  & \rxj\ \ac{nw} & 2009-01-30 &  17:11:44.4, $-$39:32:57.1 & 347.2999, $-$0.0654 &  28.4  & --\\
10091  & \rxj\ \ac{nw} & 2009-05-16 &  17:11:46.3, $-$39:32:55.7 & 347.3038, $-$0.0701 &  29.6  &  --\\
10092  & \rxj\ \ac{nw} & 2009-09-10 &  17:11:46.1, $-$39:33:51.6 & 347.3043, $-$0.0689 &  29.2  &  --\\
12671  & \rxj\ \ac{nw} & 2011-07-01 &  17:11:47.5, $-$39:33:41.2 & 347.2959, $-$0.0807 &  89.9  &  --\\
\hline
6721 & CTB~37A &   2006-10-07 &  17:14:35.8, $-$38:31:24.6 & 348.4558, $+$0.0879 &  19.9  & 70.3 \\
14586 & IC~4637 &   2014-03-07 &  17:05:10.5, $-$40:53:08.4 & 345.4794, $+$0.1403 &  29.6   & 109.8 \\
\enddata
\tablenotetext{\ast}{Galactic longitude and latitude coordinates expressed in degrees.}
\tablenotetext{\dagger}{Angular distance from \rxj\ \ac{nw} with ObsID of 12671.}
\label{tab:chandra_dataset}
\end{deluxetable*}

\section{Analysis and Results} \label{sec:AnalysisAndResults}

\subsection{Hotspot} \label{subsec:Hot-spot}

We present a flux image at 0.5--7.0 keV of the \ac{nw} shell of \rxj\  in \figref{fig:1713_regs}.
This image was generated using {\tt merge\_obs} in CIAO, which combined all the observations of \rxj\ listed in \tabref{tab:chandra_dataset}.
We found many bright, point-like sources, and conducted a detailed study to investigate the properties of these point-like structures.

\begin{figure}[thb!]
\plotone{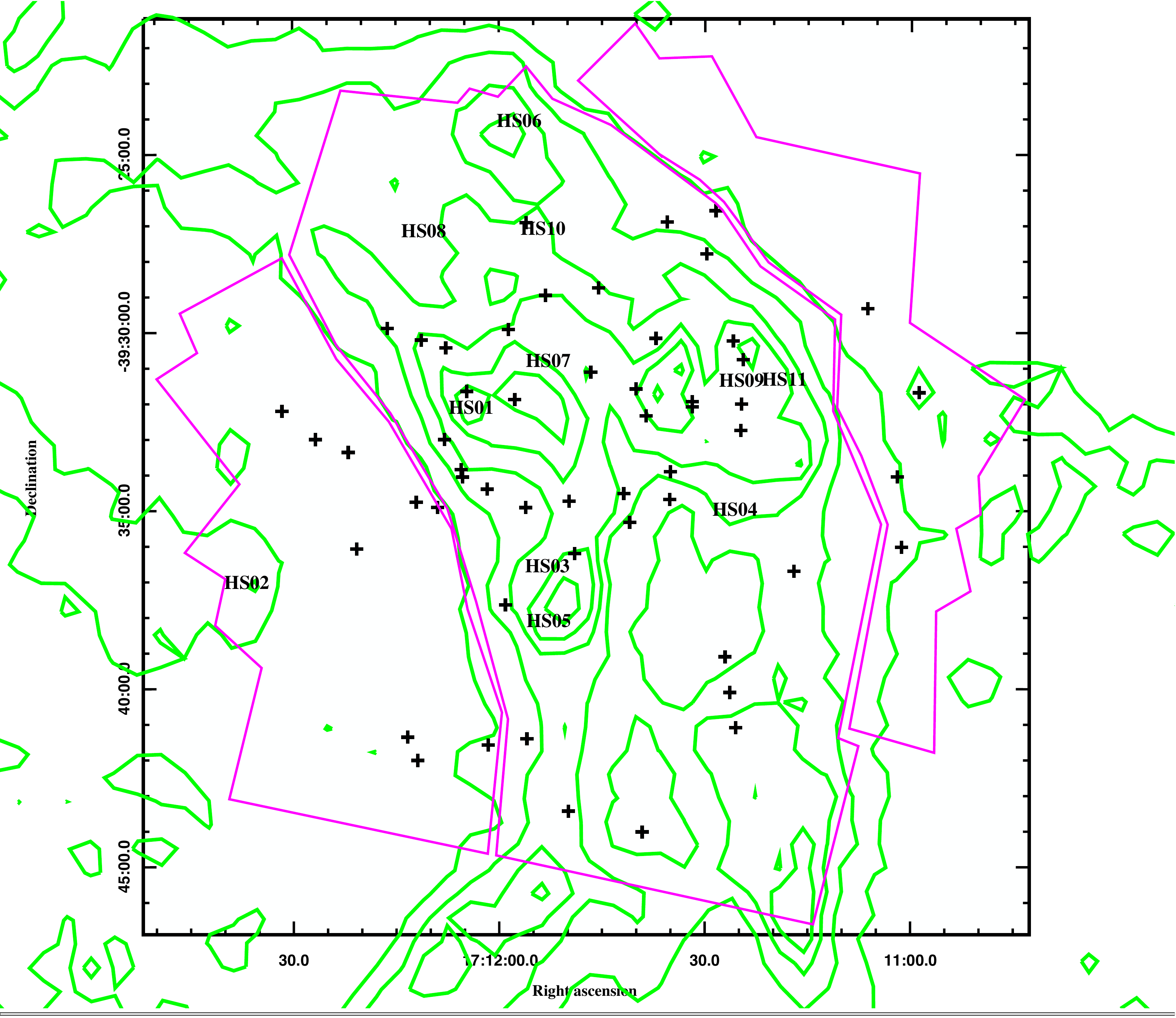}
\caption{
The locations of HS01--HS65 in the \ac{nw} rim of \rxj. HS01 to HS11 (top 11 brightest hotspots) are shown with their names, whereas the others are shown with black crosses. The contour is taken with \xmm\ (0.5--8 keV), and the magenta regions are the same regions shown in \figref{fig:1713_regs}.
}
\label{fig:HS_loc}
\end{figure}

We selected 65 sources in the following procedure and defined them as ``\hsS’’.
First, the point-like sources were systematically picked up using {\tt wavdetect} in CIAO.
In this process, we used a combined count map in the range of 0.5--7.0 keV and
a combined \ac{psf} map that is weighted by an exposure time of each epoch.
We ran {\tt wavdetect} with the ``wavelet scale’’ of 1 and 2, which is a parameter of a radius of the wavelet function (see \citet{Freeman2002a} for details).
This resulted in the detection of 154 sources.
Second, we set a criterion of the photon flux to be larger than $1.0 \times 10^{-6}\ \mathrm{photon\ cm^{-2}\ s^{-1}}$,
reducing the number of the detected sources to 65.
The detection significance of these 65 sources was more than 4$\sigma$.
Note that sources with photon fluxes smaller than the criterion value have limited statistics, causing large uncertainty in the spectral analysis.
In this study, we analyzed in detail the properties of the 65 \hsS, hereafter labeled as HS01 to HS65.
We show their locations in the \ac{nw} rim of the \ac{snr} in \figref{fig:HS_loc}.
HS01 to HS65 are labeled in the order of their X-ray brightness.
Their locations and the other properties (obtained below) are summarized in \tabref{tab:HSs_summary}.

Most of the \hsS\ were spatially consistent with the point-like sources, whereas some showed slight spatial extension.
We could not confidently test their spatial extension due to the poor statistics.

\begin{figure}[thb!]
\plotone{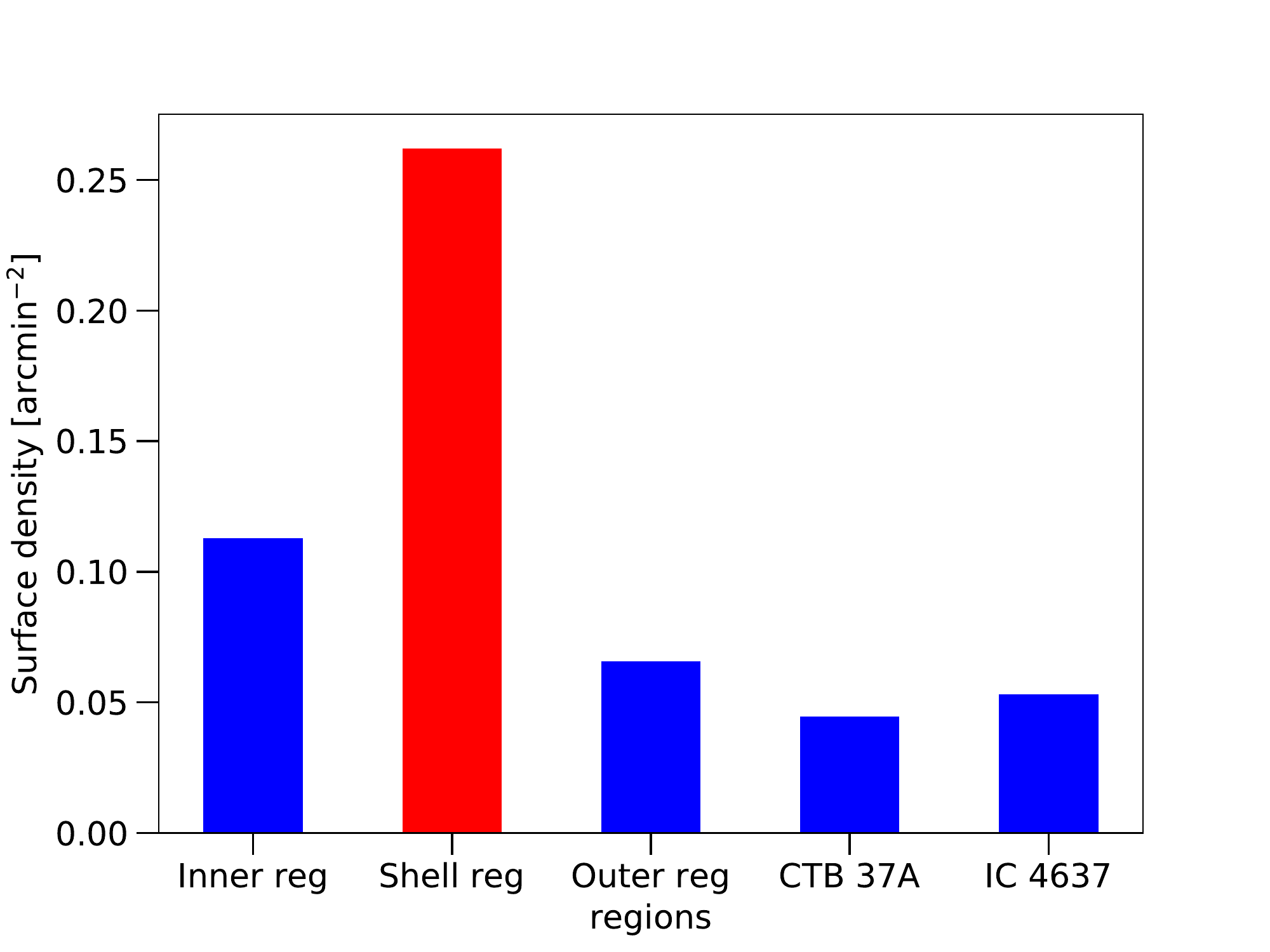}
\caption{
The surface density of \hs\ in \rxj\ \ac{nw}, CTB 37A, and IC~4637.
We divide the \ac{nw} region of \rxj\ into three regions, labeled Inner reg, Shell reg, and Outer reg, as shown in \figref{fig:1713_regs}.
}
\label{fig:HS_density}
\end{figure}

We confirmed that the \ac{nw} shell of \rxj\ contains many \hs-like structures.
\figref{fig:HS_density} illustrates the surface density of the \hs, i.e., summing the number of \hsS\ contained in ``Shell reg’’, ``Inner reg’’, and ``Outer reg’’, shown in \figref{fig:1713_regs}, divided by their areas.
The radial widths of these regions are roughly $\sim 0.2^\circ$ (Shell reg), $\sim 0.1^\circ$ (Inner reg), and $\sim 0.1^\circ$ (Outer reg), assuming the center of the \ac{snr} being ($17^\mathrm{h}13^\mathrm{m}25.2^\mathrm{s}$, $-39^\mathrm{d}46^\mathrm{m}15.6^\mathrm{s}$) \citep{H.E.S.S.Collaboration2018a}.
The azimuthal angle of these regions from the center is $\sim 60^\circ$.
There is clearly an excess number of \hsS\ in the shell,
which is one of the brightest regions in the whole \ac{snr}.
Because the \ac{nw} part of \rxj\ is located in the Galactic Plane,
many X-ray sources likely exist both in the foreground and background.
However, we found that the number of the \hs\ is quite large in \rxj\ \ac{nw}, as follows.
We applied the same method of detecting \hs-like features to the other two \chandra\ observations performed in the vicinity of \rxj, namely, SNR CTB~37A and the planetary nebula IC~4637.
They are $70.3$\arcmin\ and $109.8$\arcmin\ away from the \ac{nw} rim of \rxj, respectively (\tabref{tab:chandra_dataset}).
Note that we extracted the source-free region for the observation of CTB~37A.
The surface density of the hotspots from these two observations are also shown in \figref{fig:HS_density}, labeled as ``CTB~37A’’ and ``IC~4637’’.
The larger population of \hsS\ in the \ac{nw} shell of \rxj\ suggests that the \hsS\ could be associated with the remnant.

We estimated the number of \hsS\ in the shell of the \ac{snr}.
The average of the surface density obtained from CTB~37A and IC~4637 was $\sim0.049$ arcmin$^{-2}$, and used as background.
Then, the net surface density calculated from the two regions of ``Shell reg'' and ``Inner reg'' was estimated to be 0.167 $\mathrm{arcmin^{-2}}$.
If we assume the surface density is uniform in the entire shell, simply describes by an annulus with the inner and outer radii of 0.2\degr\ and 0.5\degr, respectively,
the number of \hsS\ in the entire shell of the \ac{snr} is estimated as $\sim 400$.

\subsection{Flux Image} \label{subsec:Fluximage}

\begin{figure}[thb!]

  \plotone{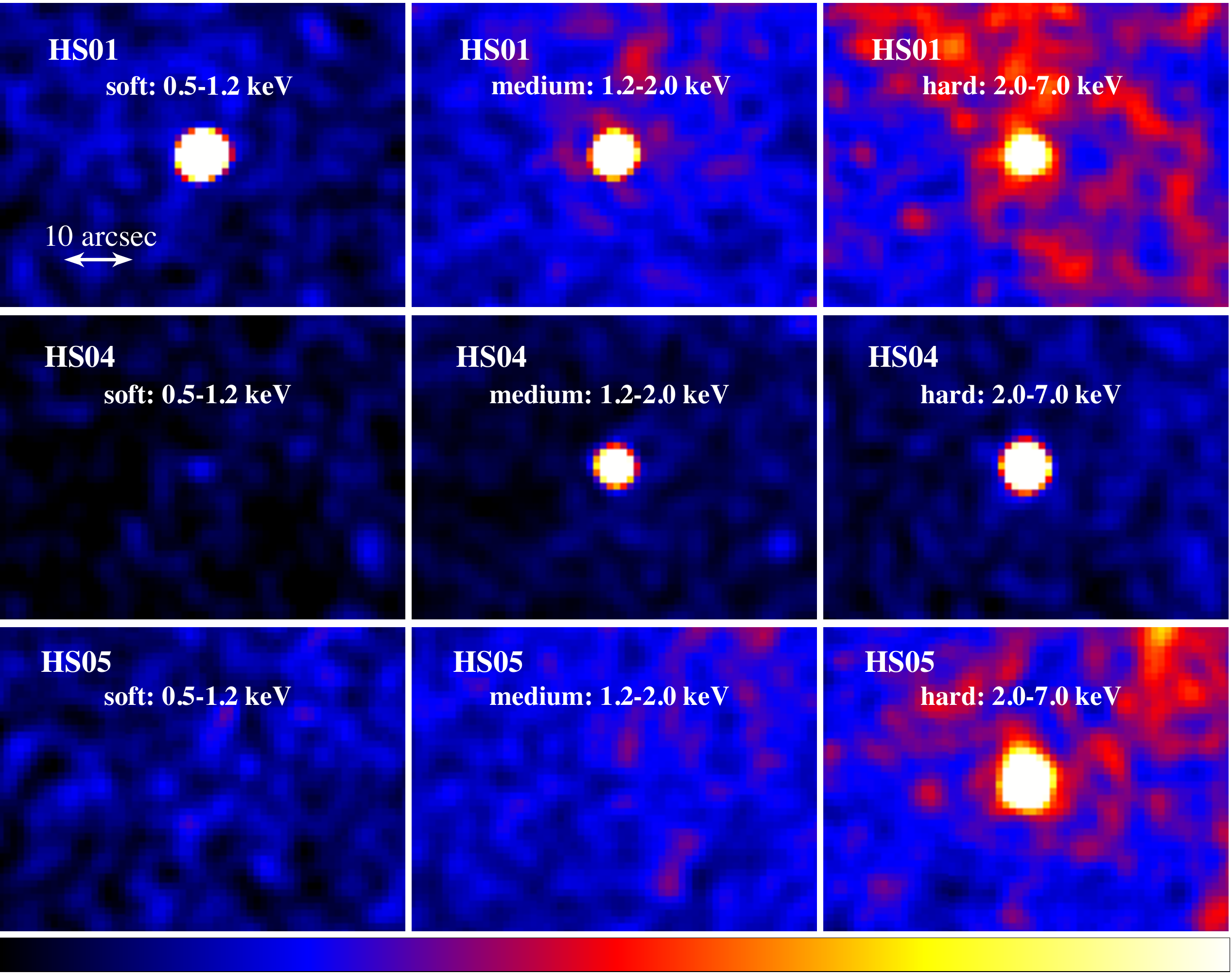}
  \caption{
   Flux images of the three hotspots (HS01, HS04, and HS05 from the top), shown in the three energy bands of $0.5\--1.2\ \mathrm{keV}$, $1.2\--2.0\ \mathrm{keV}$, and $2.0\--7.0\ \mathrm{keV}$ from the left.
  }
  \label{fig:fluximages}
\end{figure}

We produced flux images in three energy bands,
$0.5\--1.2\ \mathrm{keV}$ (soft), $1.2\--2.0\ \mathrm{keV}$ (medium), and $2.0\--7.0\ \mathrm{keV}$ (hard).
All seven epochs were combined with {\tt merge\_obs}, setting the bin size to 2 (i.e., one pixel corresponds to 1\arcsec).
The flux images of the three example \hsS\ (HS01, HS04, and HS05) are shown in \figref{fig:fluximages}.

As shown in the figure, there are various types of hotspots seen differently in the different energy channels:
HS01 is bright in the soft band, whereas HS05 is dark in the soft and medium bands.
This suggests that their spectra would also vary depending on the hotspots, which is presented in the next subsection.

\subsection{Spectrum} \label{subsec:Spectrum}

\begin{deluxetable*}{ccccccc}[ht!]
\tablecaption{ The best-fit parameters of HS01, HS04, and HS05 }
\tablehead{
\colhead{HS} & \colhead{Model\tablenotemark{$\ast$}} & \colhead{$N_H$}  & \colhead{$\Gamma/k_{\mathrm{B}}T$} & \colhead{Flux\tablenotemark{$\dagger$} (min)}   & \colhead{Flux\tablenotemark{$\dagger$} (max)} &\colhead{$\chi^2/dof$}\\
\colhead{}  &  \colhead{}      & \colhead{($10^{22}\ \mathrm{cm^{-2}}$)}  & \colhead{(/keV)} & \colhead{($10^{-14}\ \mathrm{erg\ cm^{-2}\ s^{-1}}$)} & \colhead{($10^{-14}\ \mathrm{erg\ cm^{-2}\ s^{-1}}$)} &\colhead{}
}
\startdata
\hline
HS01 & PL    & $0.423^{+0.095}_{-0.084}$ & $4.22^{+0.35}_{-0.30}$ & $2.47^{+0.45}_{-0.91}$ & $7.19^{+0.98}_{-0.94}$ &315.1/289\\
HS01 & Bremss  & 0.423 (fix)& $0.386^{+0.020}_{-0.019}$ & $2.25^{+0.58}_{-0.47}$ & $5.97^{+0.70}_{-0.63}$ &344.8/290\\
HS01 & Bbody   & 0.423 (fix)& $0.176^{+0.005}_{-0.005}$ & $2.11^{+0.52}_{-0.38}$ & $5.41^{+0.61}_{-0.68}$ & 370.1/290\\
\hline
HS04 & PL    & $4.42^{+0.49}_{-0.45}$ & $3.37^{+0.26}_{-0.24}$ & $4.22^{+0.33}_{-1.01}$ & $6.90^{+0.31}_{-1.10}$ &188.8/186\\
HS04 & Bremss  & 4.42 (fix)& $1.41^{+0.10}_{-0.09}$ & $3.92^{+0.54}_{-0.67}$ & $6.40^{+0.58}_{-0.63}$ &200.1/187\\
HS04 & Bbody   & 4.42 (fix)& $0.51^{+0.02}_{-0.02}$ & $3.50^{+0.46}_{-0.51}$ & $5.60^{+0.61}_{-0.50}$ &239.4/187\\
\hline
HS05 & PL     & $19.72^{+7.62}_{-5.94}$ & $1.02^{+0.78}_{-0.67}$ & $3.62^{+0.24}_{-1.43}$ & $23.45^{+0.41}_{-7.85}$ &145.2/172\\
HS05 & Bremss\tablenotemark{$\ddagger$}  & 19.72 (fix) & -- & -- & -- & -- \\
HS05 & Bbody   & 19.72 (fix) & $1.65^{+0.19}_{-0.15}$ & $3.42^{+0.43}_{-0.64}$ & $22.91^{+0.96}_{-2.41}$ & 145.74/173\\
\enddata
\label{tab:spectral_fit}
\tablenotetext{\ast}{PL, Bremss, and Bbody represent the models of the power law, Bremsstrahlung, and black body, respectively.}
\tablenotetext{\dagger}{Flux is calculated in the range of $0.5\--7.0\ \mathrm{keV}$}
\tablenotetext{\ddagger}{The spectrum of HS05 is not able to fit with the Bremsstrahlung model.}
\end{deluxetable*}

\begin{figure*}[thb!]

  \gridline{\fig{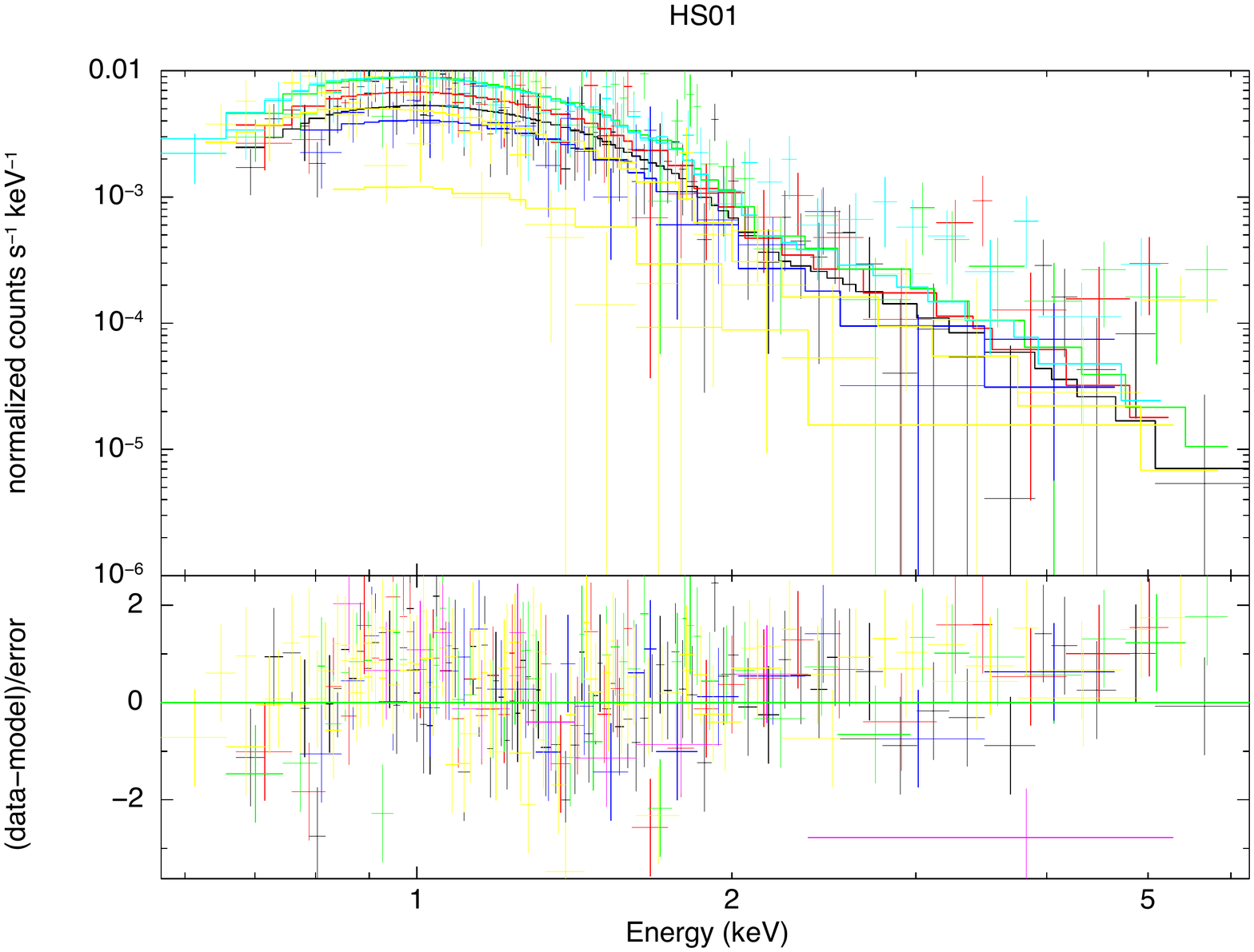}{0.3\textwidth}{HS01}
            \fig{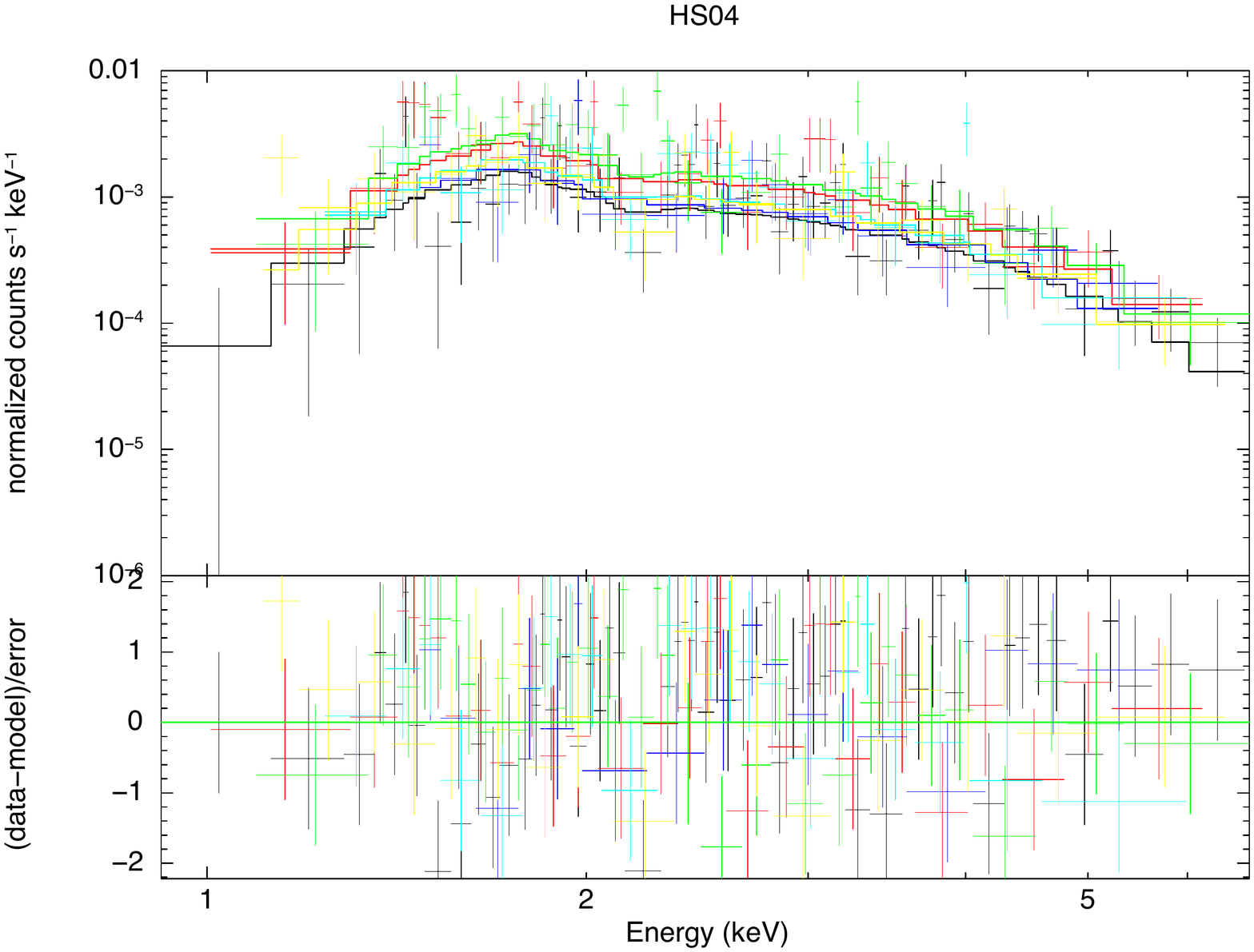}{0.3\textwidth}{HS04}
            \fig{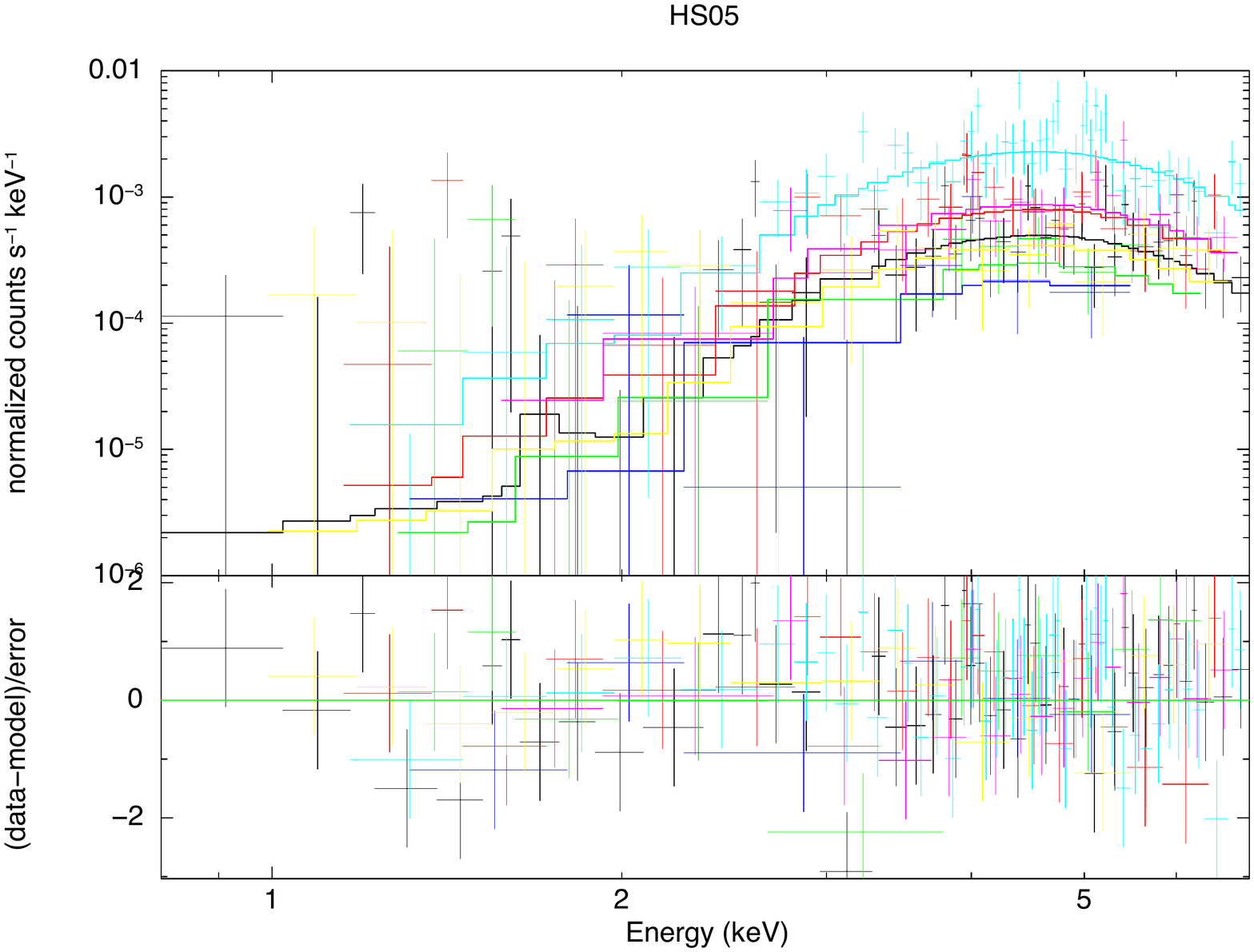}{0.3\textwidth}{HS05}            }

  \caption{
The spectrum of HS01, HS04, and HS05 from the left panel, shown with the absorbed power-law model.
The different colors represent the different epochs:
the black spectrum was taken in 2011, red in September 2009, green in May 2009, blue in January 2009, cyan in 2006, magenta in 2005, and yellow in 2000.
  }
  \label{fig:spectrum}
\end{figure*}

We performed spectral analysis of the \hs.
The spectrum of each \hs\ was extracted from a circular region with a radius of 5\arcsec\ using {\tt specextract} in CIAO.
We then subtracted the background, which was extracted from an annulus region with the inner and outer radii of 8\arcsec\ and 27\arcsec, respectively.
Both the source and background regions had the same central coordinate, which was detected by {\tt wavdetect} in \secref{subsec:Hot-spot}.
In the following spectral analysis, we used XSPEC version 12.9.1 included in HEASoft version 6.21.
The uncertainties in this paper indicate 1$\sigma$.

We used an absorbed (TBabs in XSPEC; \citet{Wilms2000a}) power-law model to fit the spectra. We jointly fitted the spectra of the different epochs:
we tied the parameters of the model (a photon index $\Gamma$ and a column density of interstellar absorption $N_H$) with the all spectra
and set the normalization free for each epoch.
\figref{fig:spectrum} presents the spectra of the three \hsS\ (HS01, HS04, and HS05), shown with the best-fit model.
In \tabref{tab:HSs_summary}, the results of each \hs\ are summarized.
The upper and lower limits were set at a 1$\sigma$ confidence level.
The flux in the energy band of 0.5$\--$7.0 keV was calculated for each observation, and the  minimum and maximum values are shown in \tabref{tab:HSs_summary}.
Furthermore, because hotspots are dark at certain epochs, spectral fit could not be well performed using all the observation data.
In this case, the fitting was performed using only epochs in which the \hs\ was bright enough, thus only the maximum value of the energy flux is shown in \tabref{tab:HSs_summary}.
It should be noted that some \hsS\ with poor statistics were not well-fitted, and their results are  shown as ``$-$'' in \tabref{tab:HSs_summary}.

\begin{figure}[thb!]
  \plotone{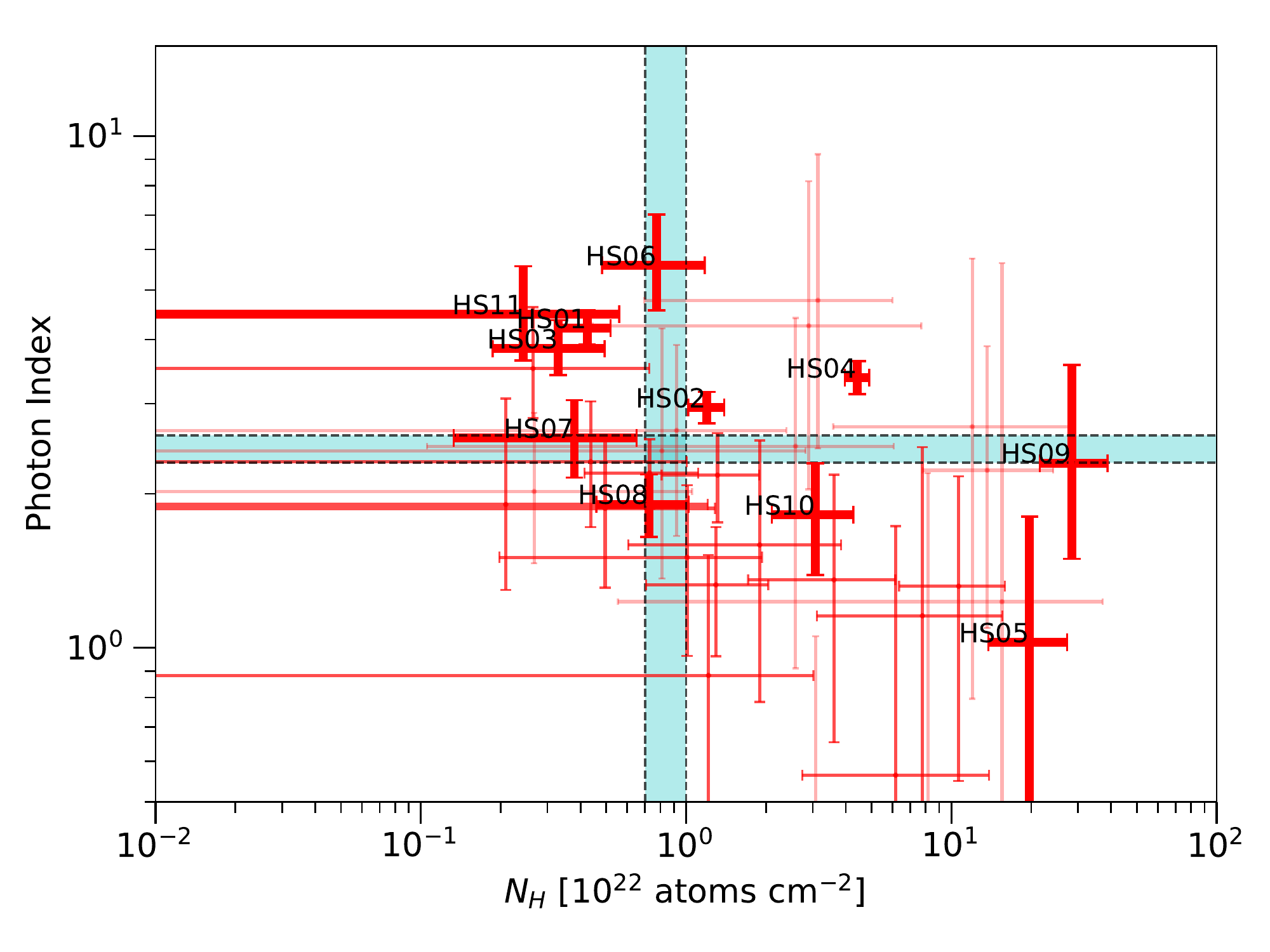}
  \caption{
 Diagram of the column density ($N_H$) and the photon index ($\Gamma$) derived from the spectral fitting of each \hs\ with the absorbed power-law model.
 Typical $N_H$ of $(0.7\--1.0)\times 10^{22}\ \mathrm{cm^{-2}}$ and $\Gamma$ of $2.3\--2.6$ in the NW shell of \rxj\ are shown in the cyan region.
 Note that the results of the bright \hs\ are highlighted with the thick red lines.
  }
  \label{fig:nH-phoIndex}
\end{figure}

In \figref{fig:nH-phoIndex}, we show the $N_H$--$\Gamma$ diagram, derived from the spectral fitting of each \hs.
Note that bright hotspots (HS01--HS11) are highlighted with red thick lines.
\figref{fig:nH-phoIndex} does not include the results of hotspots whose spectral parameters were not well constrained and were obtained only with the upper and lower limits.
The typical values of $N_H$ of $(0.7\--1.0)\times10^{22}~ \mathrm{cm}^{-2}$ and $\Gamma$ of 2.3--2.6 in \rxj\ \ac{nw} are shown with the cyan region.
As shown in \figref{fig:nH-phoIndex}, the spectral features of the hotspots substantially deviate from the typical values.
There is an anti correlation between $N_H$ and $\Gamma$: the $N_H$ of \hs\ is larger for the smaller $\Gamma$.
It should be noted that we subtracted the spectral component of the \ac{snr} as the background.
Therefore, the spectrum of the \hs\ has a pure X-ray emission from itself, and it does not contain any emission from the remnant.
We emphasize that the spectrum of the \hs\ indicates a significantly different spectrum expected for the non-thermal (synchrotron) radiation in \rxj.
The origin of these spectra is discussed in \secref{sec:Discussions}.

We also fitted the spectra of HS01, HS04, and HS05 with thermal models, Bremsstrahlung and black-body radiation.
The column density was fixed to the best-fit value derived from using the power-law model.
The results are presented in \tabref{tab:spectral_fit}.
The power-law model is favored inferred from the chi-squared value,
although the limited statistics do not confidently exclude the thermal origin.
However, the featureless spectrum and the relatively higher temperature obtained for  the thermal model might support non-thermal radiation, namely, the synchrotron emission from the accelerated electrons (see \secref{sec:Discussions} for details).

\subsection{Time Variation} \label{subsec:TimeVariation}

\begin{figure}[htb!]
\plotone{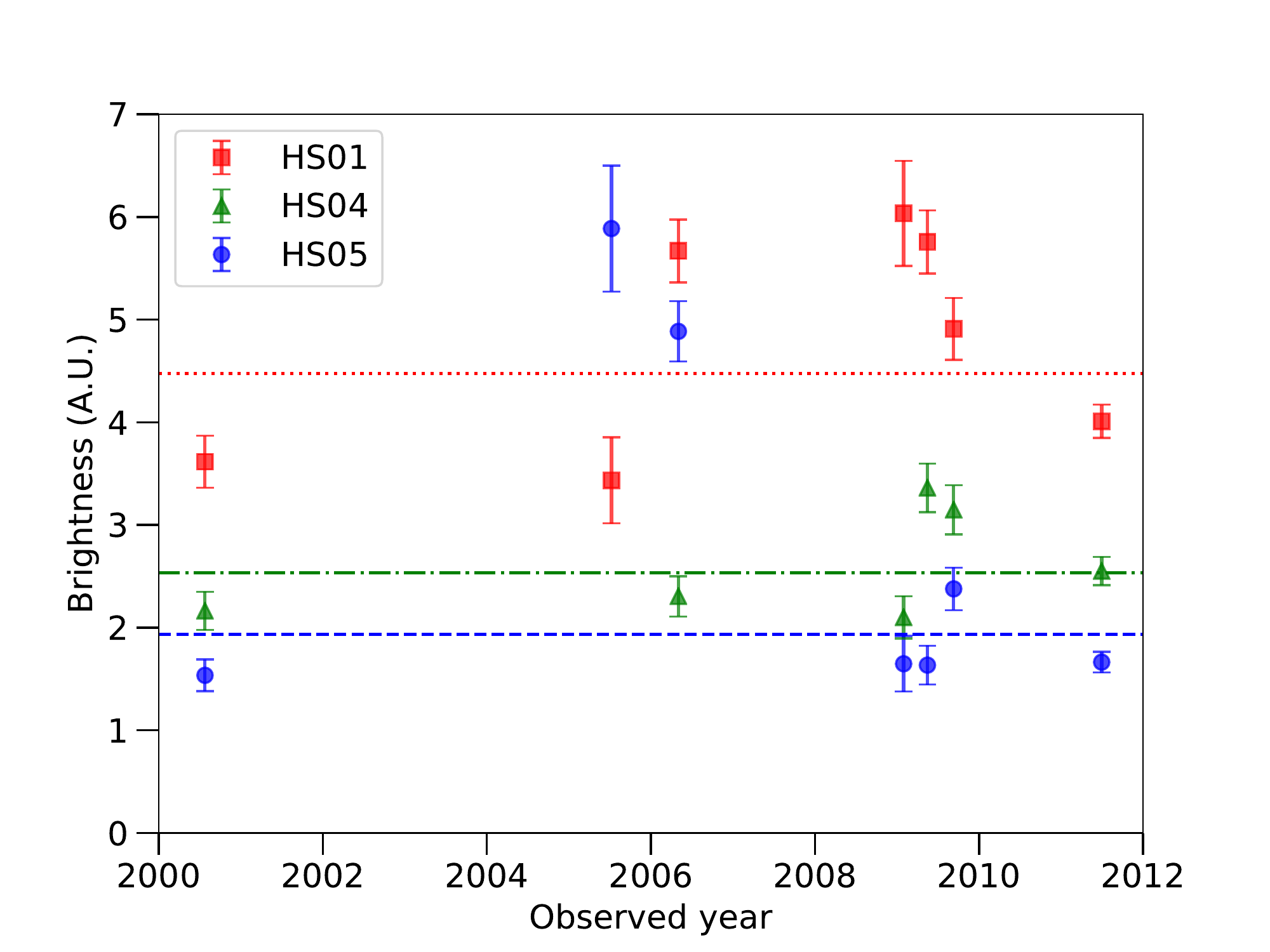}
  \caption{
  Light curves of HS01 (shown with red squares), HS04 (green triangles), and HS05 (blue circles).
The horizontal lines represent the constant values that give $\chi^2_{\rm min}$ in chi-squared test (see also the text).
  }
  \label{fig:LCs}
\end{figure}

In this section, we discuss the time variation of the \hsS.
\figref{fig:LCs} presents light curves of the three \hsS,
where the brightness was calculated by summing up the flux values inside a circular region with a radius of 5\arcsec.
The three \hsS\ show varying brightness from time to time.
We evaluated the time variability using a chi-squared ($\chi^2$) test.
The chi-squared value is given by
\begin{equation}
  \chi^2=\sum_i \frac{\left(f_i- f_c \right)^2}{\sigma_i^2} ,
  \label{eq:chi2}
\end{equation}
where $i$, $f_i$, $f_c$, and $\sigma_i$ are a subscript notating each observation, the brightness, the constant value, and the error of the brightness, respectively.
Once the minimum $\chi^2$ value ($\chi^2_{\rm min}$) is found by changing $f_c$,
we derived the confidence to exclude the hypothesis of no time variation from the value of $\chi^2_{\rm min}$.
The constant value that gives $\chi^2_{\rm min}$ is also shown in \figref{fig:LCs}.

We found that 20 \hsS\ showed time variations at more than $3\sigma$ confidence level, and 34 \hsS\ do so at more than  $2\sigma$ (see \tabref{tab:HSs_summary}).
We note that 8 \hsS\ varied not only on a year-scale but also on a month-scale with more than 3$\sigma$, inferred from the same chi-squared test using the three data taken in 2009.
The observed time variation may contain an important key for interpreting the origin of the \hsS: in fact a short time variation indicates a strong magnetic field of $\sim$mG intensity, as discussed in \citet{Uchiyama2007a}.

\section{Discussion} \label{sec:Discussions}

Our observations revealed the significantly large population of \hsS\ in \rxj\ NW,
the trend of the spectra (the larger $N_H$, the smaller $\Gamma$),
and the yearly and monthly variabilities of the flux.
We suggest that the \hsS\ are attributed to synchrotron radiation resulting in the interaction between a dense molecular core and the \ac{snr} shock.
An inhomogeneous environment, where the dense core and clumps have survived against the stellar wind from the massive progenitor star, is possible for SNR \rxj.
When the SNR shock interacts with the cores, synchrotron emission is expected due to an enhanced magnetic field surrounding the core \citep{Inoue2012a,Celli2019a}.

The timescales of synchrotron cooling and acceleration of electrons of energy $E_e$ are respectively given by
\begin{eqnarray}
  t_{\mathrm{synch}} &\approx& 12\left(\frac{B}{\mathrm{mG}}\right)^{-2}\left(\frac{E_e}{\mathrm{TeV}}\right)^{-1} \text { yr }\nonumber , \\
  &\approx& 1.5 \left(\frac{B}{\mathrm{mG}}\right)^{-3/2}\left(\frac{\varepsilon}{\mathrm{keV}}\right)^{-1/2}\ \mathrm{yr} \label{eq:cooling} ,
  \end{eqnarray}
 and
 \begin{eqnarray}
  t_{\mathrm{acc}} &\approx&  1 \eta\left(\frac{\varepsilon}{\mathrm{keV}}\right)^{1/2}\left(\frac{B}{ \mathrm{mG}}\right)^{-3/2}\left(\frac{v_{\mathrm{s}}}{ 3000\ \mathrm{km\ s^{-1}}}\right)^{-2} \text{yr} .\nonumber \\
  \label{eq:acceleration}
\end{eqnarray}
Here $B$, $\eta$, and $v_{\mathrm{s}}$ are the strength of the magnetic field, the so-called gyro factor which is defined as the mean free path of the particle divided by its gyroradius, and the shock speed, respectively.
Note that the case of $\eta=1$ is the most efficient acceleration, know as Bohm limit.
The characteristic energy of a synchrotron photon ($\varepsilon$) is given by $\varepsilon\approx 0.016 (B/\mathrm{mG})(E_{\mathrm{e}}/\mathrm{TeV})^2\ \mathrm{keV}$.

 \citet{Inoue2012a} showed that the penetration depth of a particle into a dense clump due to its random walk is described with
\begin{eqnarray}
l_{\mathrm{pd}} & \simeq&\left(\kappa_{\mathrm{d}} t\right)^{1 / 2} \nonumber , \\
   &=&0.1 \eta^{1 / 2}\left(\frac{E}{10\ \mathrm{TeV}}\right)^{1 / 2}\left(\frac{B}{100\ \mu \mathrm{G}}\right)^{-1 / 2}\left(\frac{t}{10^{3}\ \mathrm{yr}}\right)^{1 / 2} \mathrm{pc} , \nonumber \\
   \label{eq:petration}
\end{eqnarray}
where $t$ is defined as the time since the high-energy particle started penetrating into the clump.
Here, $\kappa_{\mathrm{d}}=4\eta l_{\mathrm{g}}c/3\pi$ is the diffusion coefficient\footnote{If we assume the diffusion in pitch angle is $D_\theta=\pi c/4\eta l_{\mathrm{g}}\sim c/\eta l_{\mathrm{g}}$, the diffusion coefficient becomes $\kappa_{\mathrm{d}}=\eta l_{\mathrm{g}}c/3$.} of a high-energy particle with a gyroradius of $l_{\mathrm{g}}$.

We confirmed the thermal emission from the shock-core interaction is not dominant for the case of \rxj.
The temperature of a proton in a shocked core, which corresponds to the upper limit of the temperature of electrons, is characterized by two parameters: the speed of the transmitted shock in the core and the number density ratio of the diffuse gas and the core.
The shock speed in the core is given by $v_{\mathrm{sh,c}}\approx v_{\mathrm{sh,d}}\times(n_{\mathrm{d}}/n_{\mathrm{c}})^{1/2}$, where $v_{\mathrm{sh,d}},\ n_{\mathrm{d}}$, and $n_{\mathrm{c}}$ are the shock speed in the diffuse gas, the number density of the diffuse gas, and the number density of the core, respectively.
The ambient density of \rxj\ was estimated to be $n_{\mathrm{d}}\sim 0.01\ \mathrm{cm^{-3}}$ from X-ray observations \citep{Slane1999a,Cassam-Chenai2004a,Takahashi2008a,Tanaka2008a,Katsuda2015a,Tsuji2016a}.
High-density objects such as cores and clumps are not heated enough to emit X-rays by the transmitted shock because the shock speed in the core is roughly hundreds of $\mathrm{m\ s^{-1}}$.
The timescale of plasma instability (Kelvin-Helmholtz instability) is of the same order as the so-called ``cloud-crushing time'', $t_{\mathrm{cc}}$ \citep{Klein1994a}.
Assuming a cloud with a size of $L_{\mathrm{c}}$, the cloud-crushing time is given by $t_{\mathrm{cc}}=L_{\mathrm{c}}/v_{\mathrm{sh,c}}$.
This results in tens of thousands of years, consistently with the estimate in \citet{Celli2019a}.\footnote{In fact, \citet{Celli2019a} used ``clump crossing time'', which is calculated as $2L_{\mathrm{c}}/v_{\mathrm{sh,c}}\sim1.4\times 10^{4}\ \mathrm{years}$.}
Thus, the clump survives against the plasma instability because $t_{\mathrm{cc}}$ is much longer than the age of the SNR.
Therefore, we presume that the X-ray emitted from the \hs\ is attributed to predominantly non-thermal (synchrotron) radiation.
A non-thermal origin is also supported by the featureless spectra.

We propose two scenarios for the origin of the \hs:
the X-ray emission originates from ``primary electrons’’ and/or from ``secondary electrons.’’
In this paper, primary electrons indicates those accelerated by shocks (i.e., the forward shock of the \ac{snr} and/or the reflected shock induced by the interaction between the forward shock and the core).
Secondary electrons are those produced by the decay of a charged pion, which is one of the products arising from the collision of very-high-energy protons and matters in the core.
We note that secondary positrons are also taken into consideration.

In the following subsections, we discuss the two scenarios.
Hereafter, we assume that $\eta\approx 1$ and $v_s=$ 3900 km/s in the case of \rxj\ \ac{nw} \citep{Uchiyama2007a,Tanaka2008a,Tsuji2016a,Tsuji2019a}.

\subsection{\HS\ Originated from Primary Electrons}
\label{subsec:PrimaryElectronHS}

When the SNR shock is interacting with the dense core,
the magnetic field is enhanced, resulting in the strong synchrotron emission from the electrons accelerated at the SNR shock.
As shown in \secref{subsec:TimeVariation}, 20 \hsS\ are variable in the temporal interval of a few years, and 8 \hsS\ vary even in few months, at more than 3$\sigma$ confidence levels.
The fast time variability of the synchrotron radiation might indicate the presence of amplified magnetic field.
Indeed, $B=0.5\--2.0$ mG could explain the observed year-scale and month-scale variations.
Assuming $\varepsilon=0.5\--10$ keV and $B=0.5\--2\ \mathrm{mG}$,
we obtained $ t_{\mathrm{synch}}=0.2\--6.0$ yr and $ t_{\mathrm{acc}}=0.1\--5.3$ yr from \eqref{eq:cooling} and \eqref{eq:acceleration}, respectively.

Substituting the synchrotron cooling time to \eqref{eq:petration}, the penetration depth of an electron is written as
\begin{equation}
  l_{\mathrm{pd,e}}\approx0.026 \left(\frac{B}{100\ \mathrm{\mu G}}\right)^{-3 / 2}\ \mathrm{pc}
  \label{eq:electronpenetration}
\end{equation}
 \citep{Fukui2012a}.
Note that the penetration depth is independent on the energy of the electrons.
It depends on the amplitude of the turbulent magnetic field with the scale of gyroradius of the electrons.
Using \eqref{eq:electronpenetration},
$l_\mathrm{pd,e}$ is roughly estimated to be 2.3 mpc (milli parsec) for $B$=0.5 mG,
and 0.3 mpc for $B$=2.0 mG.

The \hs\ seems to be spatially comparable with the size of \ac{psf} of \chandra,
which is approximately 1 arcsec.
This corresponds to a radius of $\sim 5\ \mathrm{mpc}$, assuming the distance to the \ac{snr} of $\sim 1\ \mathrm{kpc}$ \citep{Fukui2003a}.
Therefore, the radius of the X-ray bright region in the core is smaller than 5 mpc.

The number density is typically $\sim 10^{5} \--10^6\ \mathrm{cm^{-3}}$ in the core of  molecular clouds.
The Atacama Large Millimeter/Submillimeter Array (ALMA) lately found cores with number densities of $\sim 10^{7}\ \mathrm{cm^{-3}}$ in the regions like Taurus and Orion~A \citep[e.g.,][]{onishi2015a,ohashi2016a,ohashi2018a}.
We therefore assumed $n = 10^{5}\--10^{7}\ \mathrm{cm^{-3}}$ for the number density of the core.

\begin{figure}[ht!]
\plotone{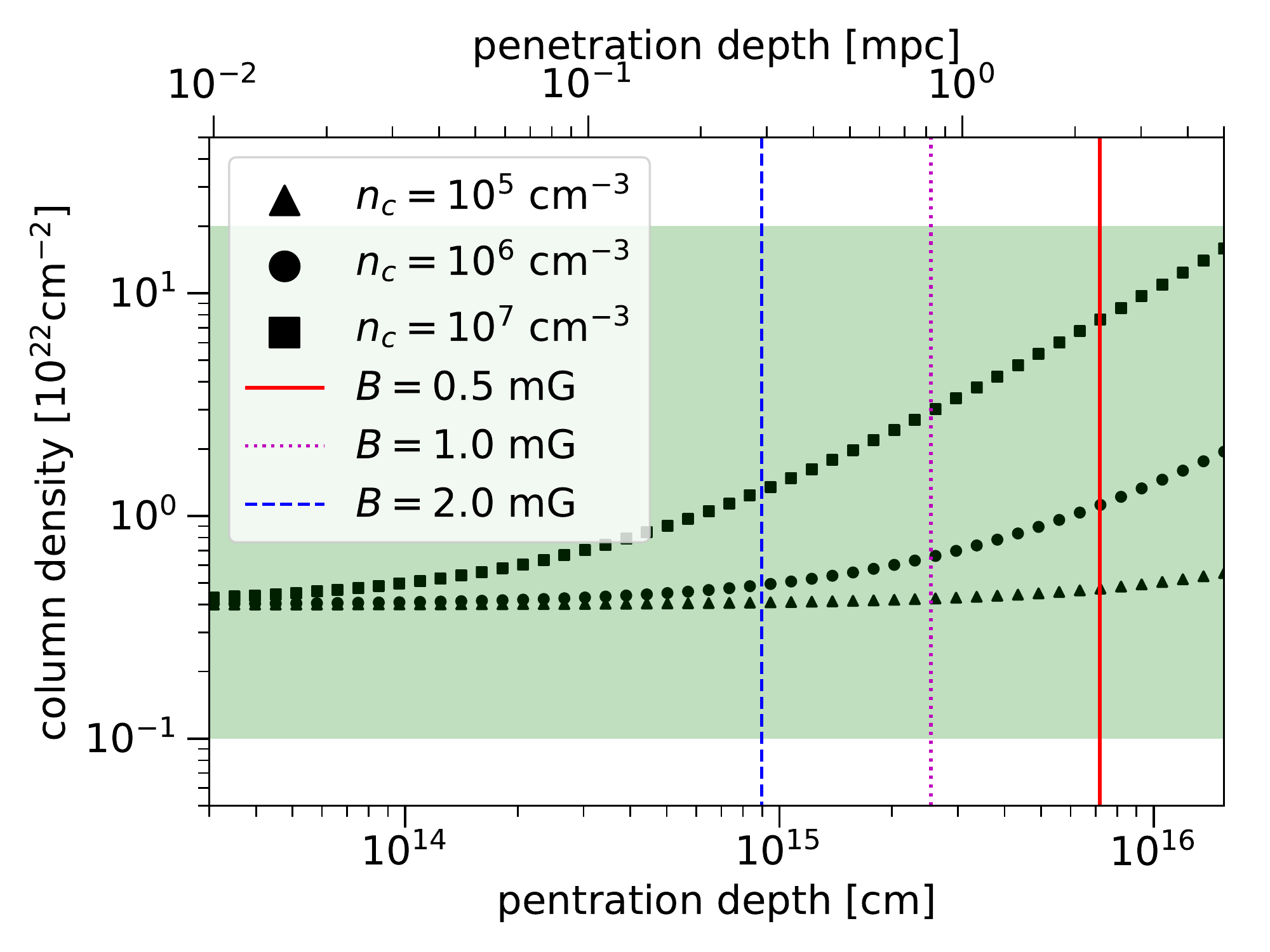}
  \caption{
The estimated column density as a function of the penetration depth,
assuming the number density of the core of $10^5,$ $10^6$, and $10^7\ \mathrm{cm^{-3}}$ and using \eqref{eq:N}.
The vertical red-solid, magenta-dotted, and blue-dashed lines respectively represent the penetration depth of an electron in magnetic fields of 0.5 mG, 1.0 mG, and 2.0 mG using \eqref{eq:electronpenetration}.
The observed column density ($N_H=(0.1\--20)\times 10^{22}\ \mathrm{cm^{-2}}$) is shown in the green region.
}
  \label{fig:columndensity}
\end{figure}

The column density 
($N$) is described as
\begin{equation}
N=n\times l_{\mathrm{pd,e}} + N_\mathrm{H,LOS},
\label{eq:N}
\end{equation}
where the first and second terms account for the column density of the core and the line-of-sight column density, respectively.
Here, $N_\mathrm{H,LOS}$ is assumed to be the column density observed in the central region of the SNR, $0.4 \times 10^{22}\ \mathrm{cm^{-2}}$ \citep{Cassam-Chenai2004a},
because the central part is assumed to be less contaminated by the swept-up ISM than the shell region, and thus expected to have the information of the wind-blown cavity.
\figref{fig:columndensity} illustrates the relation between the estimated column density ($N$) and $l_\mathrm{pd,e}$,
assuming $n_c$ of $10^{5}$, $10^{6}$, and $10^{7}$ $\mathrm{cm^{-3}}$.
In \figref{fig:columndensity}, the penetration depth of an electron into magnetic fields of $0.5\ \mathrm{mG}$, $1.0\ \mathrm{mG}$, and $2.0\ \mathrm{mG}$ are shown with the solid, dotted, and dashed vertical lines, respectively.

In the case of a strong magnetic field ($B$=0.5 mG),
$l_\mathrm{pd,e} $ was calculated to be $\sim$2.3 mpc. 
This would result in a column density of $N \sim 1 \times 10^{23}\ \mathrm{cm^{-2}}$  for $n_c = 10^7\ \mathrm{cm^{-3}}$ (\eqref{eq:N}), which is roughly consistent with the observed values.
Note that a lower column density, $N \lesssim 1 \times 10^{22}\ \mathrm{cm^{-2}}$, is obtained for $n_c = 10^{5}\--10^{6}\ \mathrm{cm^{-3}}$.

In the case of a very strong magnetic field ($B$=2 mG),
$l_\mathrm{pd,e}$ becomes $\sim$0.3 mpc, which is much smaller than the size of the core.
In this case, the electron cannot deeply penetrate to the core because it would quickly lose its energy in the skin of the core.
This reduces the column density to
$N \lesssim 1 \times 10^{22}\ \mathrm{cm^{-2}}$ in the case of $n_{\mathrm{c}}=10^ {5}\--10^{7}\ \mathrm{cm^{-3}}$.

The spectral analysis shows that most of the \hsS\ are characterised by the low $N_{\mathrm{H}}$ and large $\Gamma$ (\figref{fig:nH-phoIndex}).
This can be interpreted as acceleration in the reflection shock caused by the interaction between the forward shock and the clump.
The spectrum of electrons accelerated in the reflection shock becomes steeper due to the low Mach number of the reflection shock (see \citet{Inoue2012a} for details).

In summary,
for the \hs\ originating from primary electrons,
the observed properties of the \hs\ are interpreted as follows.
The rapid flux variation is explained by the amplified magnetic field of 0.5--2 mG.
The larger $N_H$ is interpreted as the deeper penetration depth in the  $B=0.5$ mG case,
whereas the smaller $N_H$ might be indicative of the shallower penetration depth in a very strong magnetic field of 2 mG.
Furthermore, the observed higher $\Gamma$ might be caused by the reflection shock with the lower Mach number.

\subsection{\HS\ originated from secondary electrons}
\label{subsec:SecondaryElectronHS}
The \hsS, the hard \hs\ in particular, are likely attributed to the secondary electrons that are produced via the shock-cloud interaction.
We assume the same penetration depth of the secondary electrons as that of protons.
The hardest \hs\ tends to have the largest column density, as shown in \figref{fig:nH-phoIndex}.
The spectrum of high-energy protons in the core is expected to be harder
since low-energy protons do not penetrate the clump deeply, and only very-high-energy protons are capable to reach the core \citep{Gabici2014a,Celli2019a}.
Therefore, one expects the harder spectrum of the secondary electrons produced via the shock-clump interaction, and also the harder spectrum of the synchrotron radiation from the secondary electrons, which is reconciled with the observed small $\Gamma$.

The timescale of proton-proton (pp) interaction is characterized by $t_\mathrm{pp} \sim 6 \times 10^{7}(n/1\ \mathrm{cm^{-3}})^{-1}\ \mathrm{yr}$ with the number density of the target matter of $n$.\footnote{
In the derivation of $t_\mathrm{pp}$, we used the values of $\sim40\ \mathrm{mb}$ and $0.45$ for the cross section and inelasticity, respectively \citep{Aharonian1996a}.}
In the case of $n \sim 10^7\ \mathrm{cm^{-3}}$, $t_\mathrm{pp}$ is $\sim$6 years, which is roughly consistent with the observed year-scale variability.
Furthermore, the timescale of pp interaction is less than the age of \rxj, given $n=10^5\--10^7\ \mathrm{cm^{-3}}$.
Substituting the timescale of pp interaction to \eqref{eq:petration},
we derive the penetration depth of the proton,
 \begin{eqnarray}
   &&l_{\mathrm{pd,p}}\sim\nonumber \\
   &&7.7\left(\frac{E_{\mathrm{p}}}{10\ \mathrm{TeV}}\right)^{1/2}\left(\frac{B}{100\ \mathrm{\mu G}}\right)^{-1/2}\left(\frac{n}{10^7\ \mathrm{cm^{-3}}}\right)^{-1/2}\ \mathrm{mpc},\nonumber\\
   \label{eq:lpd_p}
 \end{eqnarray}
where $E_{\mathrm{p}}$ is the energy of the proton.
Note that $l_{\mathrm{pd,p}}$ depends on the number density of the core.

In the case of the core with $n=10^7\ \mathrm{cm^{-3}}$,
an energy of about 40 TeV is required for a proton to reach the vicinity of the center of the core,
assuming $B=$1 mG and $l_{\mathrm{pd,p}} \sim$5 mpc in \eqref{eq:lpd_p}.
This results in a column density $N$ of $\sim 10^{23}\ \mathrm{cm^{-2}}$, which is roughly consistent with the observed larger $N_H$.
The energy of the proton should be less than 200 TeV, otherwise the proton passes through the core without pp interaction because of the longer penetration depth than the size of the X-ray emitting region (i.e., $l_{\mathrm{pd,p}} \geq$ 10 mpc).
Protons with energy of $\sim$40 TeV ($\leq$200 TeV) can produce the secondary electron responsible for the synchrotron radiation in the X-ray energy band.

In the case of the core with $n=10^5\ \mathrm{cm^{-3}}$,
the proton energy of about 0.4 TeV is required to reach the vicinity of the center of the core,
assuming $B=$1 mG and $l_{\mathrm{pd,p}} \sim$5 mpc in \eqref{eq:lpd_p}.
The energy of the proton should be less than 2 TeV in order to trigger the pp interact within the X-ray emitting region.
For protons with energy of $\sim$0.4 TeV ($\leq$2 TeV), the energy is too low to produce the secondary electron responsible for the synchrotron radiation in the X-ray energy band.
Therefore, the very large density of $n=10^6$--$10^7\ \mathrm{cm^{-3}}$ is preferred for the hard \hs\ in the scenario of secondary electrons.

In the following, we estimate the synchrotron energy flux emitted from secondary electrons in order to verify the consistency with the observations.
The observed flux of the hard \hs\ is $\sim 10^{-15}$--$10^{-14}\ \mathrm{erg\ cm^{-2}\ s^{-1}}$, as derived in \secref{subsec:Spectrum}.

The energy density of protons in the \ac{snr} is described by
\begin{eqnarray}
U_{\mathrm{p}}&=&\frac{K_{\mathrm{p}}}{V}, \\
K_{\mathrm{p}}&=&\xi E_{\mathrm{SN}}=\xi\times 10^{51}\ \mathrm{erg} .
\end{eqnarray}
Here $U_{\mathrm{p}}$, $K_{\mathrm{p}}$, $V$, $\xi$, and $E_{\mathrm{SN}}$ are the energy density of protons, the total energy of protons filled in the volume of the \ac{snr}, the volume of the \ac{snr}, the fraction of the kinetic energy transferred from the supernova explosion to protons with energy of 10$\--$1000 TeV, and the typical kinetic energy of supernova explosion, respectively.
Suppose that the protons have the power-law distribution of $N(E)\propto E^{-2}$, $\xi$ is derived from the fraction ($\sim$33\%) of energy occupied by the protons from 10 TeV to 1 PeV in the total energy range from 1 GeV to 1 PeV.
This results in $\xi \approx  10\% \times 33\%  \approx 0.033$.
Note that we assumed the typical fraction of the kinetic energy transferred from the supernova explosion to accelerated protons was 10\%.
The radius of the remnant is $\sim$30~arcmin, which corresponds to $\sim 9\ \mathrm{pc}$, assuming $d=1$~kpc \citep{Fukui2003a}.
If the cosmic-ray protons are isotropically distributed inside the remnant, the total energy of protons contained in the X-ray emitting region is calculated by
\begin{eqnarray}
W_{\mathrm{p}}= U_{\mathrm{p}} \times \frac{4}{3}\pi R^3 ,
\end{eqnarray}
where $R$ is the radius of the X-ray emitting region, namely the core  ($R=5\ \mathrm{mpc}$).
The synchrotron flux emitted by secondary electrons in the characteristic timescale $\tau$ is then given by
\begin{eqnarray}
F_X &=&\frac{\zeta W_{\mathrm{p}}}{4\pi d^2 \tau} \\
&\sim& 8\times10^{-20} \left(\frac{\xi}{1}\right)\left(\frac{n}{1\ \mathrm{cm^3}}\right)\ \mathrm{erg}~ \mathrm{cm}^{-2}~ \mathrm{s}^{-1} .
\label{eq:estimated_flux}
\end{eqnarray}
We assume that the fraction of the energy transferred from primary protons to secondary electrons ($\zeta$) is 0.1,
and the timescale ($\tau$) is comparable with the timescale of pp interactions, $\tau \sim 6 \times 10^{7}(n/1\ \mathrm{cm^{-3}})^{-1}\ \mathrm{yr}$.
We also assume that secondary electrons lose all their energy by synchrotron X-ray radiation in the amplified strong magnetic field.


In the case of $n=10^{6}\--10^{7}\ \mathrm{cm^{-3}}$ and $\xi=0.033$, \eqref{eq:estimated_flux} yields $F_X \sim 10^{-15}\-- 10^{-14}\ \mathrm{erg\ cm^{-2}\ s^{-1}}$, which reproduces the observed flux level.
In the case of $n=10^5\ \mathrm{cm^{-3}}$, $\xi$ should be $\sim 0.33$ to reach  $\sim 10^{15}\ \mathrm{erg\ cm^{-2}\ s^{-1}}$. 
$\xi\sim 0.33$, however, is unrealistically large since typical fraction of the kinetic energy transferred from supernova explosion to accelerated protons is $\sim 0.1$.
We also argued that $n=10^5\ \mathrm{cm^{-3}}$ is not likely applicable
because the expected energy of the proton inside the core is lower for producing the secondary electron responsible for the synchrotron X-ray.

If the \hsS\ originate from secondary electrons, one expects the hadronic gamma-ray radiation produced from the same protons.
It is challenging but very interesting to observe the hadronic gamma-ray radiation from the \hs\ and compare it with the X-ray properties.
The current gamma-ray telescopes, with the limited angular resolution, make it unrealistic to detect sub-arcsec structures of the \hsS.
\ac{cta}, however, will provide much better spatial resolution and sensitivity, and allow us to access the gamma-ray information of the hotspots.
Neutrino observations may also contribute to this study of revealing the hadronic component.

\section{Conclusions} \label{sec:conclusion}

Using \chandra\ observations of the NW rim of SNR \rxj,
we report, for the first time, the detection of a bunch of \hsS\ likely associated with the \ac{snr}.
The spectra of the \hsS\ are well described by an absorbed power-law model with widely distributed parameters of $N_H \sim 10^{21}\--10^{23}\ \mathrm{cm^{-2}}$ and $\Gamma \sim 0.5\-- 6$.
These parameters show the larger $N_H$ for the smaller $\Gamma$, and
are completely different from those in the remnant.
We also found that X-ray intensities of about one-third of the 65 \hsS\ were variable on a yearly scale, and 8 \hsS\ showed monthly variabilities.
We suggest that these \hsS\ are attributed to the synchrotron X-ray radiation resulting from the interaction between the SNR shock and the dense molecular cores.
The synchrotron radiation in the shock-core interaction originates from the accelerated electrons interacting with the core (primary electron origin) and/or electrons produced via accelerated protons colliding with the core (secondary electron origin).

\acknowledgments
Acknowledgments
We thank the anonymous referee for the useful comments.
R. H. is supported by Rikkyo University Special Fund for Research.
N. T. is supported by the Japan Society for the Promotion of Science (JSPS) KAKENHI grant No. JP17J06025.
This work was supported by KAKENHI Grant Numbers 18H03722.
Support for this work was provided by the National Aeronautics and Space Administration through Chandra Award Numbers GO9-0074X and GO1-12092X issued by the Chandra X-ray Observatory Center, which is operated by the Smithsonian Astrophysical Observatory for and on behalf of the National Aeronautics Space Administration under contract NAS8-03060.

\software{CIAO (v4.9)\citep{Fruscione2006a},
          HEASoft (v6.21)\citep{HEASARC2014a},
          XSPEC (v12.9.1)\citep{Arnaud1996a}
          }




\newpage
\begin{longrotatetable}
\begin{deluxetable*}{cccccccccc}
\tablecaption{
The summary of HS01$\--$65
}
\tablewidth{700pt}
\tabletypesize{\scriptsize}
\tablehead{
\colhead{HS ID}& \twocolhead{Position}&  \colhead{Detection significance} &\colhead{$N_H$} &\colhead{$\Gamma$} &\colhead{Flux (min)} &\colhead{Flux (max)} &\colhead{$\chi^2/dof$} &\colhead{Time variability}\\
\colhead{} & \twocolhead{($\alpha,\ \delta$)} & \colhead{($\sigma$)} & \colhead{($10^{22}\ \mathrm{cm^{-2}}$)} & \colhead{} & \colhead{($10^{-14}\ \mathrm{erg\ cm^{-2}\ s^{-1}}$)} & \colhead{($10^{-14}\ \mathrm{erg\ cm^{-2}\ s^{-1}}$)} & \colhead{} &\colhead{($\geq3\sigma$)}
}
\startdata
\label{tab:HSs_summary}
HS01 & 17:12:04.125 & -39:32:10.859 & 73.4 & $0.42_{-0.08}^{+0.10}$ & $4.22_{-0.30}^{+0.35}$ & $2.47_{-0.91}^{+0.45}$ & $7.19_{-0.94}^{+0.98}$ & 315.1/289 & Y\\
HS02 & 17:12:36.796 & -39:37:05.230 & 19.5 & $1.20_{-0.18}^{+0.20}$ & $2.95_{-0.20}^{+0.22}$ & $4.54_{-0.60}^{+0.30}$ & $4.86_{-0.43}^{+0.29}$ & 76.4/92 & N\\
HS03 & 17:11:52.997 & -39:36:38.134 & 49.5 & $0.33_{-0.14}^{+0.16}$ & $3.84_{-0.43}^{+0.53}$ & $1.17_{-0.47}^{+0.33}$ & $3.59_{-0.75}^{+0.23}$ & 155.7/157 & N\\
HS04 & 17:11:25.658 & -39:35:02.662 & 58.5 & $4.42_{-0.45}^{+0.49}$ & $3.37_{-0.24}^{+0.26}$ & $4.22_{-1.01}^{+0.33}$ & $6.90_{-1.10}^{+0.31}$ & 188.8/186 & YM\\
HS05 & 17:11:52.790 & -39:38:09.992 & 26.9 & $19.72_{-5.94}^{+7.62}$ & $1.02_{-0.67}^{+0.78}$ & $3.62_{-1.43}^{+0.24}$ & $23.45_{-7.85}^{+0.41}$ & 145.2/172 & Y\\
HS06 & 17:11:57.055 & -39:24:07.441 & 14.1 & $0.77_{-0.29}^{+0.40}$ & $5.59_{-1.03}^{+1.44}$ & $1.82_{-0.89}^{+0.05}$ & $2.30_{-1.10}^{+0.21}$ & 67.3/79 & N\\
HS07 & 17:11:52.863 & -39:30:51.863 & 25.7 & $0.38_{-0.25}^{+0.27}$ & $2.57_{-0.42}^{+0.48}$ & $0.52_{-0.23}^{+0.16}$ & $1.83_{-0.49}^{+0.31}$ & 87.6/109 & YM\\
HS08 & 17:12:10.965 & -39:27:13.396 & 19.0 & $0.72_{-0.26}^{+0.30}$ & $1.91_{-0.26}^{+0.28}$ & $0.85_{-0.29}^{+0.23}$ & $3.39_{-0.44}^{+0.28}$ & 77.3/98 & YM\\
HS09 & 17:11:24.721 & -39:31:25.237 & 25.7 & $28.41_{-6.85}^{+10.38}$ & $2.30_{-0.80}^{+1.27}$ & $\leq 2.82$ & $\leq 6.31$ & 105.2/112 & N\\
HS10 & 17:11:53.593 & -39:27:09.311 & 17.6 & $3.07_{-0.97}^{+1.20}$ & $1.82_{-0.43}^{+0.47}$ & $1.00_{-0.53}^{+0.24}$ & $2.47_{-0.78}^{+0.09}$ & 68.6/77 & Y\\
HS11 & 17:11:18.509 & -39:31:23.079 & 15.6 & $\leq 0.56$ & $4.49_{-0.84}^{+1.08}$ & $1.16_{-0.99}^{+0.18}$ & $2.51_{-1.72}^{+0.15}$ & 71.7/96 & Y\\
HS12 & 17:11:28.422 & -39:26:34.798 & 16.6 & $10.65_{-4.30}^{+5.28}$ & $1.32_{-0.77}^{+0.84}$ & $2.50_{-1.48}^{+0.05}$ & $4.84_{-2.73}^{+0.30}$ & 46.2/56 & N\\
HS13 & 17:12:21.943 & -39:33:21.784 & 17.7 & $1.31_{-0.50}^{+0.57}$ & $2.17_{-0.42}^{+0.45}$ & $0.67_{-0.25}^{+0.25}$ & $2.90_{-0.58}^{+0.44}$ & 42.7/62 & Y\\
HS14 & 17:11:40.029 & -39:31:35.142 & 15.2 & $\leq 0.86$ & $3.41_{-0.77}^{+1.42}$ & $0.38_{-0.37}^{+0.15}$ & $1.06_{-0.85}^{+0.26}$ & 55.4/86 & N\\
HS15 & 17:11:35.522 & -39:26:53.704 & 15.2 & $\leq 0.73$ & $3.51_{-0.70}^{+1.12}$ & $0.57_{-0.41}^{+0.08}$ & $1.59_{-0.88}^{+0.04}$ & 50.8/52 & N\\
HS16 & 17:12:07.920 & -39:33:00.300 & 13.1 & $\leq 0.52$ & $2.29_{-0.26}^{+0.66}$ & $0.43_{-0.36}^{+0.29}$ & $2.73_{-1.48}^{+0.08}$ & 63.6/75 & YM\\
HS17 & 17:11:53.219 & -39:28:57.455 & 13.6 & $3.61_{-1.90}^{+2.53}$ & $1.36_{-0.70}^{+0.82}$ & $\leq 0.52$ & $2.92_{-1.16}^{+0.46}$ & 66.8/82 & Y\\
HS18 & 17:12:11.875 & -39:42:00.805 & 9.1 & $\leq 0.44$ & $1.39_{-0.35}^{+0.42}$ & $0.42_{-0.39}^{+0.25}$ & $2.09_{-1.02}^{+0.04}$ & 16.7/13 & N\\
HS19 & 17:11:01.348 & -39:36:01.349 & 11.6 & $1.29_{-0.59}^{+0.75}$ & $1.33_{-0.37}^{+0.40}$ & $\--$ & $8.13_{-1.17}^{+0.42}$ & 43.7/42 & YM\\
HS20 & 17:11:45.483 & -39:28:45.032 & 14.0 & $6.16_{-3.43}^{+7.70}$ & $0.56_{-0.78}^{+1.17}$ & $0.94_{-0.72}^{+0.54}$ & $2.90_{-1.14}^{+0.27}$ & 42.4/59 & Y\\
HS21 & 17:11:24.699 & -39:32:00.496 & 15.6 & $\leq 1.29$ & $1.87_{-0.56}^{+0.69}$ & $0.67_{-0.43}^{+0.32}$ & $1.60_{-0.78}^{+0.29}$ & 64.9/69 & N\\
HS22 & 17:11:55.942 & -39:41:24.307 & 14.0 & $0.73_{-0.32}^{+0.38}$ & $2.20_{-0.32}^{+0.36}$ & $\leq 0.27$ & $3.93_{-0.59}^{+0.41}$ & 42.1/50 & Y\\
HS23 & 17:11:57.698 & -39:31:53.047 & 9.0 & $\leq 1.01$ & $2.31_{-0.59}^{+0.72}$ & $\leq 0.38$ & $1.39_{-0.60}^{+0.27}$ & 87.6/104 & Y\\
HS24 & 17:11:26.349 & -39:40:06.366 & 11.6 & $1.01_{-0.81}^{+0.92}$ & $1.50_{-0.54}^{+0.58}$ & $0.72_{-0.24}^{+0.11}$ & $1.85_{-0.63}^{+0.37}$ & 34.2/52 & N\\
HS25 & 17:12:01.715 & -39:34:23.946 & 11.4 & $7.78_{-4.67}^{+7.79}$ & $1.16_{-1.05}^{+1.31}$ & $\leq 0.54$ & $3.34_{-2.71}^{+0.62}$ & 52.5/64 & N\\
HS26 & 17:11:49.890 & -39:43:26.415 & 7.1 & $\leq 3.02$ & $0.88_{-0.57}^{+0.63}$ & $\leq 0.47$ & $4.71_{-1.23}^{+0.31}$ & 26.8/29 & N\\
HS27 & 17:11:31.848 & -39:31:56.608 & 9.6 & $1.89_{-1.29}^{+1.95}$ & $1.59_{-0.81}^{+0.95}$ & $\leq 0.35$ & $1.51_{-0.92}^{+0.12}$ & 58.7/67 & N\\
HS28 & 17:11:24.786 & -39:32:44.837 & 10.9 & $\leq 1.21$ & $1.91_{-0.61}^{+1.16}$ & $\leq 0.20$ & $1.16_{-0.70}^{+0.18}$ & 34.0/49 & N\\
HS29 & 17:11:39.132 & -39:44:01.427 & 6.0 & $\leq 0.34$ & $2.30_{-0.38}^{+0.86}$ & $0.74_{-0.47}^{+0.03}$ & $0.78_{-0.52}^{+0.01}$ & 11.0/15 & N\\
HS30 & 17:12:20.736 & -39:36:04.401 & 9.6 & $\leq 0.40$ & $2.84_{-0.40}^{+1.13}$ & $\leq 0.19$ & $1.39_{-1.28}^{+0.60}$ & 28.6/24 & Y\\
HS31 & 17:11:48.976 & -39:36:12.435 & 10.0 & $13.65_{-5.85}^{+10.55}$ & $2.22_{-1.13}^{+1.66}$ & $\leq 0.48$ & $\leq 2.93$ & 55.9/70 & N\\
HS32 & 17:11:46.629 & -39:31:06.732 & 9.3 & $\--$ & $\--$ & $\--$ & $\--$ & $\--$/$\--$ & N\\
HS33 & 17:12:16.244 & -39:29:52.913 & 6.8 & $\leq 2.82$ & $2.43_{-1.06}^{+1.78}$ & $\leq 0.35$ & $\leq 1.38$ & 36.0/42 & N\\
HS34 & 17:12:11.292 & -39:30:12.754 & 6.9 & $\leq 0.43$ & $2.01_{-0.30}^{+0.68}$ & $\leq 0.12$ & $1.05_{-0.47}^{+0.00}$ & 63.2/72 & Y\\
HS35 & 17:11:41.824 & -39:34:31.392 & 8.8 & $\leq 1.05$ & $2.02_{-0.56}^{+0.86}$ & $\leq 0.25$ & $1.55_{-0.83}^{+0.05}$ & 43.6/50 & N\\
HS36 & 17:11:35.140 & -39:34:41.255 & 11.2 & $\leq 9.95$ & $-0.37_{-1.19}^{+1.42}$ & $0.30_{-0.28}^{+0.25}$ & $3.42_{-1.79}^{+0.98}$ & 29.0/31 & N\\
HS37 & 17:12:01.602 & -39:41:35.049 & 7.4 & $\leq 2.39$ & $2.66_{-1.01}^{+1.25}$ & $\leq 0.36$ & $\leq 1.89$ & 19.4/23 & N\\
HS38 & 17:11:24.442 & -39:30:45.768 & 7.8 & $7.51_{-6.19}^{+4.17}$ & $\geq 3.01$ & $\leq 0.02$ & $\leq 0.73$ & 44.7/53 & N\\
HS39 & 17:12:05.302 & -39:34:03.442 & 8.0 & $\leq 1.28$ & $2.59_{-0.62}^{+2.15}$ & $\leq 0.12$ & $0.52_{-0.51}^{+0.01}$ & 34.7/42 & N\\
HS40 & 17:11:35.042 & -39:33:54.616 & 9.3 & $15.51_{-14.96}^{+21.63}$ & $1.23_{-2.53}^{+4.41}$ & $\leq 0.55$ & $\leq 1.63$ & 31.8/39 & N\\
HS41 & 17:11:25.888 & -39:30:14.223 & 6.9 & $2.90_{-2.46}^{+4.80}$ & $4.26_{-2.22}^{+3.90}$ & $\leq 0.11$ & $\leq 0.74$ & 43.5/49 & N\\
HS42 & 17:11:31.863 & -39:32:05.315 & 7.4 & $\leq1.99$ & $\geq 2.88$ & $\leq 0.06$ & $\leq 0.10$ & 51.2/50 & N\\
HS43 & 17:11:06.326 & -39:29:19.369 & 7.8 & $\leq 1.30$ & $1.10_{-0.31}^{+0.84}$ & $0.97_{-0.46}^{+0.04}$ & $1.34_{-0.66}^{+0.13}$ & 10.4/11 & N\\
HS44 & 17:12:13.337 & -39:41:21.576 & 4.9 & $\leq 6.98$ & $1.28_{-0.79}^{+4.62}$ & $\--$ & $\leq 1.10$ & 2.8/2 & N\\
HS45 & 17:12:04.689 & -39:31:39.247 & 4.8 & $\leq 2.22$ & $0.58_{-0.80}^{+2.78}$ & $\leq 0.26$ & $0.99_{-0.89}^{+0.48}$ & 67.4/66 & N\\
HS46 & 17:11:56.101 & -39:34:54.946 & 7.8 & $\--$ & $\--$ & $\--$ & $\--$ & $\--$/$\--$ & N\\
HS47 & 17:11:56.054 & -39:26:54.958 & 6.7 & $2.59_{-2.48}^{+3.47}$ & $2.48_{-1.56}^{+1.93}$ & $\leq 0.29$ & $\leq 0.93$ & 22.6/32 & N\\
HS48 & 17:11:25.488 & -39:41:05.711 & 5.6 & $8.15_{-7.02}^{+16.92}$ & $-0.01_{-1.75}^{+2.21}$ & $0.60_{-0.58}^{+0.05}$ & $1.98_{-1.91}^{+0.20}$ & 32.3/29 & N\\
HS49 & 17:11:17.037 & -39:36:41.864 & 7.9 & $\leq 4.56$ & $3.07_{-1.35}^{+1.79}$ & $\leq 0.35$ & $\leq 1.72$ & 10.8/24 & NM\\
HS50 & 17:12:31.583 & -39:32:12.191 & 4.3 & $\--$ & $\--$ & $\--$ & $\--$ & $\--$/$\--$ & N\\
HS51 & 17:12:26.721 & -39:33:00.127 & 5.9 & $\leq 0.47$ & $3.04_{-0.54}^{+1.00}$ & $\leq 0.33$ & $\leq 0.71$ & 13.3/20 & N\\
HS52 & 17:11:27.062 & -39:39:06.295 & 7.2 & $11.99_{-8.40}^{+16.59}$ & $2.70_{-1.91}^{+3.06}$ & $\leq 0.40$ & $\leq 2.27$ & 16.0/20 & N\\
HS53 & 17:11:58.633 & -39:29:54.987 & 5.2 & $10.84_{-8.51}^{+28.82}$ & $\geq 0.16$ & $\leq 0.45$ & $\leq 8.29$ & 38.3/49 & N\\
HS54 & 17:12:05.491 & -39:33:50.944 & 6.7 & $\leq 0.16$ & $1.86_{-0.42}^{+0.59}$ & $0.37_{-0.27}^{+0.11}$ & $0.79_{-0.49}^{+0.18}$ & 44.5/58 & N\\
HS55 & 17:11:38.575 & -39:32:20.384 & 5.7 & $\--$ & $\--$ & $\--$ & $\--$ & $\--$/$\--$ & YM\\
HS56 & 17:12:08.933 & -39:34:54.495 & 7.3 & $\--$ & $\--$ & $\--$ & $\--$ & $\--$/$\--$ & N\\
HS57 & 17:11:37.148 & -39:30:09.914 & 7.4 & $\leq 29.63$ & $\geq -2.98$ & $\leq 0.00$ & $\leq 6.06$ & 29.4/33 & N\\
HS58 & 17:11:49.802 & -39:34:44.236 & 9.0 & $\--$ & $\--$ & $\--$ & $\--$ & $\--$/$\--$  & N\\
HS59 & 17:11:59.065 & -39:37:38.978 & 5.8 & $\leq 0.47$ & $1.60_{-0.54}^{+0.75}$ & $\leq 0.16$ & $0.50_{-0.32}^{+0.04}$ & 16.6/22 & N\\
HS60 & 17:10:58.832 & -39:31:40.962 & 6.6 & $\--$ & $\--$ & $\--$ & $\--$ & $\--$/$\--$ & N\\
HS61 & 17:11:29.741 & -39:27:47.529 & 6.6 & $3.14_{-2.44}^{+2.86}$ & $4.77_{-2.32}^{+4.44}$ & $\leq 0.02$ & $\leq 0.39$ & 20.5/22 & N\\
HS62 & 17:12:07.730 & -39:30:25.835 & 5.5 & $\leq 16.01$ & $\geq 4.47$ & $\leq 0.22$ & $\leq 1.15$ & 39.7/46 & N\\
HS63 & 17:12:12.049 & -39:34:45.676 & 7.5 & $\leq 0.62$ & $2.40_{-0.77}^{+1.81}$ & $\--$ & $\leq 0.37$ & 9.8/9 & N\\
HS64 & 17:11:40.961 & -39:35:19.848 & 6.3 & $10.29_{-7.68}^{+10.36}$ & $\geq 2.19$ & $\--$ & $\leq 0.23$ & 12.3/15 & YM\\
HS65 & 17:11:01.986 & -39:34:02.649 & 6.5 & $\leq 3.21$ & $3.27_{-1.53}^{+3.45}$ & $\--$ & $\leq 0.32$ & 7.2/5 & N\\
\enddata
\tablecomments{
In the columns of column density($N_H$), photon index($\Gamma$) and energy flux (min and max), the errors and upper/lower limits indicate $1\sigma$ confidence.
In the column of time variability, Y and M represent detection of time variation with more than $3\sigma$ on a year scale and a month scale, respectively, and N indicates non detection.
}
\end{deluxetable*}
\end{longrotatetable}

\newpage
\bibliography{higurashi_hot-spot_v9.5.bib}

\begin{thebibliography}{}
\expandafter\ifx\csname natexlab\endcsname\relax\def\natexlab#1{#1}\fi
\providecommand{\url}[1]{\href{#1}{#1}}
\providecommand{\dodoi}[1]{doi:~\href{http://doi.org/#1}{\nolinkurl{#1}}}
\providecommand{\doeprint}[1]{\href{http://ascl.net/#1}{\nolinkurl{http://ascl.net/#1}}}
\providecommand{\doarXiv}[1]{\href{https://arxiv.org/abs/#1}{\nolinkurl{https://arxiv.org/abs/#1}}}

\bibitem[{{Abdo} {et~al.}(2011){Abdo}, {Ackermann}, {Ajello}, {Allafort},
  {Baldini}, {Ballet}, {Barbiellini}, {Baring}, {Bastieri}, {Bellazzini},
  {Berenji}, {Blandford}, {Bloom}, {Bonamente}, {Borgland}, {Bouvier},
  {Brandt}, {Bregeon}, {Brigida}, {Bruel}, {Buehler}, {Buson}, {Caliandro},
  {Cameron}, {Caraveo}, {Casandjian}, {Cecchi}, {Chaty}, {Chekhtman}, {Cheung},
  {Chiang}, {Cillis}, {Ciprini}, {Claus}, {Cohen-Tanugi}, {Conrad}, {Corbel},
  {Cutini}, {de Angelis}, {de Palma}, {Dermer}, {Digel}, {Silva}, {Drell},
  {Drlica-Wagner}, {Dubois}, {Dumora}, {Favuzzi}, {Ferrara}, {Fortin},
  {Frailis}, {Fukazawa}, {Fukui}, {Funk}, {Fusco}, {Gargano}, {Gasparrini},
  {Gehrels}, {Germani}, {Giglietto}, {Giordano}, {Giroletti}, {Glanzman},
  {Godfrey}, {Grenier}, {Grondin}, {Guiriec}, {Hadasch}, {Hanabata}, {Harding},
  {Hayashida}, {Hayashi}, {Hays}, {Horan}, {Jackson}, {J{\'o}hannesson},
  {Johnson}, {Kamae}, {Katagiri}, {Kataoka}, {Kerr}, {Kn{\"o}dlseder}, {Kuss},
  {Lande}, {Latronico}, {Lee}, {Lemoine-Goumard}, {Longo}, {Loparco},
  {Lovellette}, {Lubrano}, {Madejski}, {Makeev}, {Mazziotta}, {McEnery},
  {Michelson}, {Mignani}, {Mitthumsiri}, {Mizuno}, {Moiseev}, {Monte},
  {Monzani}, {Morselli}, {Moskalenko}, {Murgia}, {Naumann-Godo}, {Nolan},
  {Norris}, {Nuss}, {Ohsugi}, {Okumura}, {Orlando}, {Ormes}, {Paneque},
  {Parent}, {Pelassa}, {Pesce-Rollins}, {Pierbattista}, {Piron}, {Pohl},
  {Porter}, {Rain{\`o}}, {Rando}, {Razzano}, {Reimer}, {Reposeur}, {Ritz},
  {Romani}, {Roth}, {Sadrozinski}, {Saz Parkinson}, {Sgr{\`o}}, {Smith},
  {Smith}, {Spandre}, {Spinelli}, {Strickman}, {Tajima}, {Takahashi},
  {Takahashi}, {Tanaka}, {Thayer}, {Thayer}, {Thompson}, {Tibaldo}, {Tibolla},
  {Torres}, {Tosti}, {Tramacere}, {Troja}, {Uchiyama}, {Vandenbroucke},
  {Vasileiou}, {Vianello}, {Vilchez}, {Vitale}, {Waite}, {Wang}, {Winer},
  {Wood}, {Yamamoto}, {Yamazaki}, {Yang}, \& {Ziegler}}]{Abdo2011a}
{Abdo}, A.~A., {Ackermann}, M., {Ajello}, M., {et~al.} 2011, \apj, 734, 28,
  \dodoi{10.1088/0004-637X/734/1/28}

\bibitem[{{Acero} {et~al.}(2017{\natexlab{a}}){Acero}, {Katsuda}, {Ballet}, \&
  {Petre}}]{Acero2017a}
{Acero}, F., {Katsuda}, S., {Ballet}, J., \& {Petre}, R. 2017{\natexlab{a}},
  \aap, 597, A106, \dodoi{10.1051/0004-6361/201629618}

\bibitem[{{Acero} {et~al.}(2017{\natexlab{b}}){Acero}, {Aloisio}, {Amans},
  {Amato}, {Antonelli}, {Aramo}, {Armstrong}, {Arqueros}, {Asano}, {Ashley}, \&
  et~al.}]{Acero2017b}
{Acero}, F., {Aloisio}, R., {Amans}, J., {et~al.} 2017{\natexlab{b}}, \apj,
  840, 74, \dodoi{10.3847/1538-4357/aa6d67}

\bibitem[{{Ackermann} {et~al.}(2013){Ackermann}, {Ajello}, {Allafort},
  {Baldini}, {Ballet}, {Barbiellini}, {Baring}, {Bastieri}, {Bechtol},
  {Bellazzini}, {Bland ford}, {Bloom}, {Bonamente}, {Borgland }, {Bottacini},
  {Brandt}, {Bregeon}, {Brigida}, {Bruel}, {Buehler}, {Busetto}, {Buson},
  {Caliandro}, {Cameron}, {Caraveo}, {Casandjian}, {Cecchi}, {{\c{C}}elik},
  {Charles}, {Chaty}, {Chaves}, {Chekhtman}, {Cheung}, {Chiang}, {Chiaro},
  {Cillis}, {Ciprini}, {Claus}, {Cohen-Tanugi}, {Cominsky}, {Conrad}, {Corbel},
  {Cutini}, {D'Ammando}, {de Angelis}, {de Palma}, {Dermer}, {do Couto e
  Silva}, {Drell}, {Drlica-Wagner}, {Falletti}, {Favuzzi}, {Ferrara},
  {Franckowiak}, {Fukazawa}, {Funk}, {Fusco}, {Gargano}, {Germani},
  {Giglietto}, {Giommi}, {Giordano}, {Giroletti}, {Glanzman}, {Godfrey},
  {Grenier}, {Grondin}, {Grove}, {Guiriec}, {Hadasch}, {Hanabata}, {Harding},
  {Hayashida}, {Hayashi}, {Hays}, {Hewitt}, {Hill}, {Hughes}, {Jackson},
  {Jogler}, {J{\'o}hannesson}, {Johnson}, {Kamae}, {Kataoka}, {Katsuta},
  {Kn{\"o}dlseder}, {Kuss}, {Lande}, {Larsson}, {Latronico}, {Lemoine-Goumard},
  {Longo}, {Loparco}, {Lovellette}, {Lubrano}, {Madejski}, {Massaro}, {Mayer},
  {Mazziotta}, {McEnery}, {Mehault}, {Michelson}, {Mignani}, {Mitthumsiri},
  {Mizuno}, {Moiseev}, {Monzani}, {Morselli}, {Moskalenko}, {Murgia},
  {Nakamori}, {Nemmen}, {Nuss}, {Ohno}, {Ohsugi}, {Omodei}, {Orienti},
  {Orlando}, {Ormes}, {Paneque}, {Perkins}, {Pesce-Rollins}, {Piron}, {Pivato},
  {Rain{\`o}}, {Rando}, {Razzano}, {Razzaque}, {Reimer}, {Reimer}, {Ritz},
  {Romoli}, {S{\'a}nchez-Conde}, {Schulz}, {Sgr{\`o}}, {Simeon}, {Siskind},
  {Smith}, {Spand re}, {Spinelli}, {Stecker}, {Strong}, {Suson}, {Tajima},
  {Takahashi}, {Takahashi}, {Tanaka}, {Thayer}, {Thayer}, {Thompson},
  {Thorsett}, {Tibaldo}, {Tibolla}, {Tinivella}, {Troja}, {Uchiyama}, {Usher},
  {Vandenbroucke}, {Vasileiou}, {Vianello}, {Vitale}, {Waite}, {Werner},
  {Winer}, {Wood}, {Wood}, {Yamazaki}, {Yang}, \& {Zimmer}}]{Ackermann2013a}
{Ackermann}, M., {Ajello}, M., {Allafort}, A., {et~al.} 2013, Science, 339,
  807, \dodoi{10.1126/science.1231160}

\bibitem[{{Aharonian} {et~al.}(2006){Aharonian}, {Akhperjanian}, {Bazer-Bachi},
  {Beilicke}, {Benbow}, {Berge}, {Bernl{\"o}hr}, {Boisson}, {Bolz}, {Borrel},
  {Braun}, {Breitling}, {Brown}, {Chadwick}, {Chounet}, {Cornils},
  {Costamante}, {Degrange}, {Dickinson}, {Djannati-Ata{\"\i}}, {O'C.~Drury},
  {Dubus}, {Emmanoulopoulos}, {Espigat}, {Feinstein}, {Fontaine}, {Fuchs},
  {Funk}, {Gallant}, {Giebels}, {Glicenstein}, {Goret}, {Hadjichristidis},
  {Hauser}, {Hauser}, {Heinzelmann}, {Henri}, {Hermann}, {Hinton}, {Hofmann},
  {Holleran}, {Horns}, {Jacholkowska}, {de Jager}, {Kh{\'e}lifi}, {Klages},
  {Komin}, {Konopelko}, {Latham}, {Le Gallou}, {Lemi{\`e}re},
  {Lemoine-Goumard}, {Lohse}, {Martin}, {Martineau-Huynh}, {Marcowith},
  {Masterson}, {McComb}, {de Naurois}, {Nedbal}, {Nolan}, {Noutsos}, {Orford},
  {Osborne}, {Ouchrif}, {Panter}, {Pelletier}, {Pita}, {P{\"u}hlhofer},
  {Punch}, {Raubenheimer}, {Raue}, {Rayner}, {Reimer}, {Reimer}, {Ripken},
  {Rob}, {Rolland}, {Rowell}, {Sahakian}, {Saug{\'e}}, {Schlenker},
  {Schlickeiser}, {Schuster}, {Schwanke}, {Siewert}, {Sol}, {Spangler},
  {Steenkamp}, {Stegmann}, {Superina}, {Tavernet}, {Terrier}, {Th{\'e}oret},
  {Tluczykont}, {van Eldik}, {Vasileiadis}, {Venter}, {Vincent}, {V{\"o}lk}, \&
  {Wagner}}]{Aharonian2006a}
{Aharonian}, F., {Akhperjanian}, A.~G., {Bazer-Bachi}, A.~R., {et~al.} 2006,
  \aap, 449, 223, \dodoi{10.1051/0004-6361:20054279}

\bibitem[{{Aharonian} {et~al.}(2007){Aharonian}, {Akhperjanian}, {Bazer-Bachi},
  {Beilicke}, {Benbow}, {Berge}, {Bernl{\"o}hr}, {Boisson}, {Bolz}, {Borrel},
  {Braun}, {Brion}, {Brown}, {B{\"u}hler}, {B{\"u}sching}, {Carrigan},
  {Chadwick}, {Chounet}, {Coignet}, {Cornils}, {Costamante}, {Degrange},
  {Dickinson}, {Djannati-Ata{\"\i}}, {O'C.~Drury}, {Dubus}, {Egberts},
  {Emmanoulopoulos}, {Espigat}, {Feinstein}, {Ferrero}, {Fiasson}, {Fontaine},
  {Funk}, {Funk}, {F{\"u}{\ss}ling}, {Gallant}, {Giebels}, {Glicenstein},
  {Gl{\"u}ck}, {Goret}, {Hadjichristidis}, {Hauser}, {Hauser}, {Heinzelmann},
  {Henri}, {Hermann}, {Hinton}, {Hoffmann}, {Hofmann}, {Holleran}, {Hoppe},
  {Horns}, {Jacholkowska}, {de Jager}, {Kendziorra}, {Kerschhaggl},
  {Kh{\'e}lifi}, {Komin}, {Konopelko}, {Kosack}, {Lamanna}, {Latham}, {Le
  Gallou}, {Lemi{\`e}re}, {Lemoine-Goumard}, {Lohse}, {Martin},
  {Martineau-Huynh}, {Marcowith}, {Masterson}, {Maurin}, {McComb}, {Moulin},
  {de Naurois}, {Nedbal}, {Nolan}, {Noutsos}, {Olive}, {Orford}, {Osborne},
  {Panter}, {Pelletier}, {Pita}, {P{\"u}hlhofer}, {Punch}, {Ranchon},
  {Raubenheimer}, {Raue}, {Rayner}, {Reimer}, {Reimer}, {Ripken}, {Rob},
  {Rolland}, {Rosier-Lees}, {Rowell}, {Sahakian}, {Santangelo}, {Saug{\'e}},
  {Schlenker}, {Schlickeiser}, {Schr{\"o}der}, {Schwanke}, {Schwarzburg},
  {Schwemmer}, {Shalchi}, {Sol}, {Spangler}, {Spanier}, {Steenkamp},
  {Stegmann}, {Superina}, {Tam}, {Tavernet}, {Terrier}, {Tluczykont}, {van
  Eldik}, {Vasileiadis}, {Venter}, {Vialle}, {Vincent}, {V{\"o}lk}, {Wagner},
  \& {Ward}}]{Aharonian2007a}
---. 2007, \aap, 464, 235, \dodoi{10.1051/0004-6361:20066381}

\bibitem[{{Aharonian} \& {Atoyan}(1996)}]{Aharonian1996a}
{Aharonian}, F.~A., \& {Atoyan}, A.~M. 1996, \aap, 309, 917

\bibitem[{{Ambrogi} {et~al.}(2018){Ambrogi}, {Celli}, \&
  {Aharonian}}]{Ambrogi2018a}
{Ambrogi}, L., {Celli}, S., \& {Aharonian}, F. 2018, Astroparticle Physics,
  100, 69, \dodoi{10.1016/j.astropartphys.2018.03.001}

\bibitem[{{Arnaud}(1996)}]{Arnaud1996a}
{Arnaud}, K.~A. 1996, Astronomical Society of the Pacific Conference Series,
  Vol. 101, {XSPEC: The First Ten Years}, ed. G.~H. {Jacoby} \& J.~{Barnes}, 17

\bibitem[{{Axford} {et~al.}(1977){Axford}, {Leer}, \& {Skadron}}]{Axford1977a}
{Axford}, W.~I., {Leer}, E., \& {Skadron}, G. 1977, in International Cosmic Ray
  Conference, Vol.~11, International Cosmic Ray Conference, 132

\bibitem[{{Bell}(1978)}]{Bell1978b}
{Bell}, A.~R. 1978, \mnras, 182, 147, \dodoi{10.1093/mnras/182.2.147}

\bibitem[{{Blandford} \& {Ostriker}(1978)}]{Blandford1978a}
{Blandford}, R.~D., \& {Ostriker}, J.~P. 1978, \apjl, 221, L29,
  \dodoi{10.1086/182658}

\bibitem[{{Cassam-Chena{\"\i}} {et~al.}(2004){Cassam-Chena{\"\i}},
  {Decourchelle}, {Ballet}, {Sauvageot}, {Dubner}, \&
  {Giacani}}]{Cassam-Chenai2004a}
{Cassam-Chena{\"\i}}, G., {Decourchelle}, A., {Ballet}, J., {et~al.} 2004,
  \aap, 427, 199, \dodoi{10.1051/0004-6361:20041154}

\bibitem[{{Celli} {et~al.}(2019){Celli}, {Morlino}, {Gabici}, \&
  {Aharonian}}]{Celli2019a}
{Celli}, S., {Morlino}, G., {Gabici}, S., \& {Aharonian}, F.~A. 2019, \mnras,
  487, 3199, \dodoi{10.1093/mnras/stz1425}

\bibitem[{{Ellison} {et~al.}(2010){Ellison}, {Patnaude}, {Slane}, \&
  {Raymond}}]{Ellison2010a}
{Ellison}, D.~C., {Patnaude}, D.~J., {Slane}, P., \& {Raymond}, J. 2010, \apj,
  712, 287, \dodoi{10.1088/0004-637X/712/1/287}

\bibitem[{{Fesen} {et~al.}(2012){Fesen}, {Kremer}, {Patnaude}, \&
  {Milisavljevic}}]{Fesen2012a}
{Fesen}, R.~A., {Kremer}, R., {Patnaude}, D., \& {Milisavljevic}, D. 2012, \aj,
  143, 27, \dodoi{10.1088/0004-6256/143/2/27}

\bibitem[{{Freeman} {et~al.}(2002){Freeman}, {Kashyap}, {Rosner}, \&
  {Lamb}}]{Freeman2002a}
{Freeman}, P.~E., {Kashyap}, V., {Rosner}, R., \& {Lamb}, D.~Q. 2002, \apjs,
  138, 185, \dodoi{10.1086/324017}

\bibitem[{{Fruscione} {et~al.}(2006){Fruscione}, {McDowell}, {Allen},
  {Brickhouse}, {Burke}, {Davis}, {Durham}, {Elvis}, {Galle}, {Harris},
  {Huenemoerder}, {Houck}, {Ishibashi}, {Karovska}, {Nicastro}, {Noble},
  {Nowak}, {Primini}, {Siemiginowska}, {Smith}, \& {Wise}}]{Fruscione2006a}
{Fruscione}, A., {McDowell}, J.~C., {Allen}, G.~E., {et~al.} 2006, Society of
  Photo-Optical Instrumentation Engineers (SPIE) Conference Series, Vol. 6270,
  {CIAO: Chandra's data analysis system}, 62701V

\bibitem[{{Fukui} {et~al.}(2003){Fukui}, {Moriguchi}, {Tamura}, {Yamamoto},
  {Tawara}, {Mizuno}, {Onishi}, {Mizuno}, {Uchiyama}, {Hiraga}, {Takahashi},
  {Yamashita}, \& {Ikeuchi}}]{Fukui2003a}
{Fukui}, Y., {Moriguchi}, Y., {Tamura}, K., {et~al.} 2003, \pasj, 55, L61,
  \dodoi{10.1093/pasj/55.5.L61}

\bibitem[{{Fukui} {et~al.}(2012){Fukui}, {Sano}, {Sato}, {Torii}, {Horachi},
  {Hayakawa}, {McClure-Griffiths}, {Rowell}, {Inoue}, {Inutsuka}, {Kawamura},
  {Yamamoto}, {Okuda}, {Mizuno}, {Onishi}, {Mizuno}, \& {Ogawa}}]{Fukui2012a}
{Fukui}, Y., {Sano}, H., {Sato}, J., {et~al.} 2012, \apj, 746, 82,
  \dodoi{10.1088/0004-637X/746/1/82}

\bibitem[{{Funk}(2015)}]{Funk2015a}
{Funk}, S. 2015, Annual Review of Nuclear and Particle Science, 65, 245,
  \dodoi{10.1146/annurev-nucl-102014-022036}

\bibitem[{{Gabici} \& {Aharonian}(2014)}]{Gabici2014a}
{Gabici}, S., \& {Aharonian}, F.~A. 2014, \mnras, 445, L70,
  \dodoi{10.1093/mnrasl/slu132}

\bibitem[{{H.E.S.S.~Collaboration} {et~al.}(2018){H.E.S.S.~Collaboration},
  {Abdalla}, {Abramowski}, {Aharonian}, {Ait Benkhali}, {Akhperjanian},
  {Andersson}, {Ang{\"u}ner}, {Arrieta}, {Aubert}, \&
  et~al.}]{H.E.S.S.Collaboration2018a}
{H.E.S.S.~Collaboration}, {Abdalla}, H., {Abramowski}, A., {et~al.} 2018, \aap,
  612, A6, \dodoi{10.1051/0004-6361/201629790}

\bibitem[{{Hiraga} {et~al.}(2005){Hiraga}, {Uchiyama}, {Takahashi}, \&
  {Aharonian}}]{Hiraga2005a}
{Hiraga}, J.~S., {Uchiyama}, Y., {Takahashi}, T., \& {Aharonian}, F.~A. 2005,
  \aap, 431, 953, \dodoi{10.1051/0004-6361:20047015}

\bibitem[{{Huang} {et~al.}(2018){Huang}, {Li}, {Wang}, \& {Zhao}}]{Huang2018a}
{Huang}, Y., {Li}, Z., {Wang}, W., \& {Zhao}, X. 2018, arXiv e-prints.
\newblock \doarXiv{1807.11239}

\bibitem[{{Inoue} {et~al.}(2012){Inoue}, {Yamazaki}, {Inutsuka}, \&
  {Fukui}}]{Inoue2012a}
{Inoue}, T., {Yamazaki}, R., {Inutsuka}, S.-i., \& {Fukui}, Y. 2012, \apj, 744,
  71, \dodoi{10.1088/0004-637X/744/1/71}

\bibitem[{{Jogler} \& {Funk}(2016)}]{Jogler2016a}
{Jogler}, T., \& {Funk}, S. 2016, \apj, 816, 100,
  \dodoi{10.3847/0004-637X/816/2/100}

\bibitem[{{Katsuda} {et~al.}(2015){Katsuda}, {Acero}, {Tominaga}, {Fukui},
  {Hiraga}, {Koyama}, {Lee}, {Mori}, {Nagataki}, {Ohira}, {Petre}, {Sano},
  {Takeuchi}, {Tamagawa}, {Tsuji}, {Tsunemi}, \& {Uchiyama}}]{Katsuda2015a}
{Katsuda}, S., {Acero}, F., {Tominaga}, N., {et~al.} 2015, \apj, 814, 29,
  \dodoi{10.1088/0004-637X/814/1/29}

\bibitem[{{Klein} {et~al.}(1994){Klein}, {McKee}, \& {Colella}}]{Klein1994a}
{Klein}, R.~I., {McKee}, C.~F., \& {Colella}, P. 1994, \apj, 420, 213,
  \dodoi{10.1086/173554}

\bibitem[{{Koyama} {et~al.}(1997){Koyama}, {Kinugasa}, {Matsuzaki},
  {Nishiuchi}, {Sugizaki}, {Torii}, {Yamauchi}, \& {Aschenbach}}]{Koyama1997a}
{Koyama}, K., {Kinugasa}, K., {Matsuzaki}, K., {et~al.} 1997, \pasj, 49, L7,
  \dodoi{10.1093/pasj/49.3.L7}

\bibitem[{{Koyama} {et~al.}(1995){Koyama}, {Petre}, {Gotthelf}, {Hwang},
  {Matsuura}, {Ozaki}, \& {Holt}}]{Koyama1995a}
{Koyama}, K., {Petre}, R., {Gotthelf}, E.~V., {et~al.} 1995, \nat, 378, 255,
  \dodoi{10.1038/378255a0}

\bibitem[{{Krymskii}(1977)}]{Krymskii1977a}
{Krymskii}, G.~F. 1977, Akademiia Nauk SSSR Doklady, 234, 1306

\bibitem[{{Lazendic} {et~al.}(2004){Lazendic}, {Slane}, {Gaensler}, {Reynolds},
  {Plucinsky}, \& {Hughes}}]{Lazendic2004a}
{Lazendic}, J.~S., {Slane}, P.~O., {Gaensler}, B.~M., {et~al.} 2004, \apj, 602,
  271, \dodoi{10.1086/380956}

\bibitem[{{Moriguchi} {et~al.}(2005){Moriguchi}, {Tamura}, {Tawara}, {Sasago},
  {Yamaoka}, {Onishi}, \& {Fukui}}]{Moriguchi2005a}
{Moriguchi}, Y., {Tamura}, K., {Tawara}, Y., {et~al.} 2005, \apj, 631, 947,
  \dodoi{10.1086/432653}

\bibitem[{{Nasa High Energy Astrophysics Science Archive Research Center
  (Heasarc)}(2014)}]{HEASARC2014a}
{Nasa High Energy Astrophysics Science Archive Research Center (Heasarc)}.
  2014, {HEAsoft: Unified Release of FTOOLS and XANADU}.
\newblock \doeprint{1408.004}

\bibitem[{{Ohashi} {et~al.}(2016){Ohashi}, {Sanhueza}, {Chen}, {Zhang},
  {Busquet}, {Nakamura}, {Palau}, \& {Tatematsu}}]{ohashi2016a}
{Ohashi}, S., {Sanhueza}, P., {Chen}, H.-R.~V., {et~al.} 2016, \apj, 833, 209,
  \dodoi{10.3847/1538-4357/833/2/209}

\bibitem[{{Ohashi} {et~al.}(2018){Ohashi}, {Sanhueza}, {Sakai}, {Kandori},
  {Choi}, {Hirota}, {Nguy{\#7877}n-Lu'o'ng}, \& {Tatematsu}}]{ohashi2018a}
{Ohashi}, S., {Sanhueza}, P., {Sakai}, N., {et~al.} 2018, \apj, 856, 147,
  \dodoi{10.3847/1538-4357/aab3d0}

\bibitem[{{Okuno} {et~al.}(2018){Okuno}, {Tanaka}, {Uchida}, {Matsumura}, \&
  {Tsuru}}]{Okuno2018a}
{Okuno}, T., {Tanaka}, T., {Uchida}, H., {Matsumura}, H., \& {Tsuru}, T.~G.
  2018, \pasj, 70, 77, \dodoi{10.1093/pasj/psy072}

\bibitem[{{Onishi} {et~al.}(2015){Onishi}, {Tokuda}, {Saigo}, {Kawamura},
  {Fukui}, {Matsumoto}, {Inutsuka}, {Machida}, {Tomida}, \&
  {Tachihara}}]{onishi2015a}
{Onishi}, T., {Tokuda}, K., {Saigo}, K., {et~al.} 2015, in Astronomical Society
  of the Pacific Conference Series, Vol. 499, Revolution in Astronomy with
  ALMA: The Third Year, ed. D.~{Iono}, K.~{Tatematsu}, A.~{Wootten}, \&
  L.~{Testi}, 211

\bibitem[{{Pfeffermann} \& {Aschenbach}(1996)}]{Pfeffermann1996a}
{Pfeffermann}, E., \& {Aschenbach}, B. 1996, in Roentgenstrahlung from the
  Universe, ed. H.~U. {Zimmermann}, J.~{Tr{\"u}mper}, \& H.~{Yorke}, 267--268

\bibitem[{{Sano} {et~al.}(2010){Sano}, {Sato}, {Horachi}, {Moribe}, {Yamamoto},
  {Hayakawa}, {Torii}, {Kawamura}, {Okuda}, {Mizuno}, {Onishi}, {Maezawa},
  {Inoue}, {Inutsuka}, {Tanaka}, {Matsumoto}, {Mizuno}, {Ogawa}, {Stutzki},
  {Bertoldi}, {Anderl}, {Bronfman}, {Koo}, {Burton}, {Benz}, \&
  {Fukui}}]{Sano2010a}
{Sano}, H., {Sato}, J., {Horachi}, H., {et~al.} 2010, \apj, 724, 59,
  \dodoi{10.1088/0004-637X/724/1/59}

\bibitem[{{Sano} {et~al.}(2013){Sano}, {Tanaka}, {Torii}, {Fukuda}, {Yoshiike},
  {Sato}, {Horachi}, {Kuwahara}, {Hayakawa}, {Matsumoto}, {Inoue}, {Yamazaki},
  {Inutsuka}, {Kawamura}, {Tachihara}, {Yamamoto}, {Okuda}, {Mizuno}, {Onishi},
  {Mizuno}, \& {Fukui}}]{Sano2013a}
{Sano}, H., {Tanaka}, T., {Torii}, K., {et~al.} 2013, \apj, 778, 59,
  \dodoi{10.1088/0004-637X/778/1/59}

\bibitem[{{Sano} {et~al.}(2015){Sano}, {Fukuda}, {Yoshiike}, {Sato}, {Horachi},
  {Kuwahara}, {Torii}, {Hayakawa}, {Tanaka}, {Matsumoto}, {Inoue}, {Yamazaki},
  {Inutsuka}, {Kawamura}, {Yamamoto}, {Okuda}, {Tachihara}, {Mizuno}, {Onishi},
  {Mizuno}, {Acero}, \& {Fukui}}]{Sano2015a}
{Sano}, H., {Fukuda}, T., {Yoshiike}, S., {et~al.} 2015, \apj, 799, 175,
  \dodoi{10.1088/0004-637X/799/2/175}

\bibitem[{{Slane} {et~al.}(1999){Slane}, {Gaensler}, {Dame}, {Hughes},
  {Plucinsky}, \& {Green}}]{Slane1999a}
{Slane}, P., {Gaensler}, B.~M., {Dame}, T.~M., {et~al.} 1999, \apj, 525, 357,
  \dodoi{10.1086/307893}

\bibitem[{{Takahashi} {et~al.}(2008){Takahashi}, {Tanaka}, {Uchiyama},
  {Hiraga}, {Nakazawa}, {Watanabe}, {Bamba}, {Hughes}, {Katagiri}, {Kataoka},
  {Kokubun}, {Koyama}, {Mori}, {Petre}, {Takahashi}, \&
  {Tsuboi}}]{Takahashi2008a}
{Takahashi}, T., {Tanaka}, T., {Uchiyama}, Y., {et~al.} 2008, \pasj, 60, S131,
  \dodoi{10.1093/pasj/60.sp1.S131}

\bibitem[{{Tanaka} {et~al.}(2008){Tanaka}, {Uchiyama}, {Aharonian},
  {Takahashi}, {Bamba}, {Hiraga}, {Kataoka}, {Kishishita}, {Kokubun}, {Mori},
  {Nakazawa}, {Petre}, {Tajima}, \& {Watanabe}}]{Tanaka2008a}
{Tanaka}, T., {Uchiyama}, Y., {Aharonian}, F.~A., {et~al.} 2008, \apj, 685,
  988, \dodoi{10.1086/591020}

\bibitem[{{Tsuji} \& {Uchiyama}(2016)}]{Tsuji2016a}
{Tsuji}, N., \& {Uchiyama}, Y. 2016, \pasj, 68, 108,
  \dodoi{10.1093/pasj/psw102}

\bibitem[{{Tsuji} {et~al.}(2019){Tsuji}, {Uchiyama}, {Aharonian}, {Berge},
  {Higurashi}, {Krivonos}, \& {Tanaka}}]{Tsuji2019a}
{Tsuji}, N., {Uchiyama}, Y., {Aharonian}, F., {et~al.} 2019, \apj, 877, 96,
  \dodoi{10.3847/1538-4357/ab1b29}

\bibitem[{{Uchiyama} {et~al.}(2003){Uchiyama}, {Aharonian}, \&
  {Takahashi}}]{Uchiyama2003a}
{Uchiyama}, Y., {Aharonian}, F.~A., \& {Takahashi}, T. 2003, \aap, 400, 567,
  \dodoi{10.1051/0004-6361:20021824}

\bibitem[{{Uchiyama} {et~al.}(2007){Uchiyama}, {Aharonian}, {Tanaka},
  {Takahashi}, \& {Maeda}}]{Uchiyama2007a}
{Uchiyama}, Y., {Aharonian}, F.~A., {Tanaka}, T., {Takahashi}, T., \& {Maeda},
  Y. 2007, \nat, 449, 576, \dodoi{10.1038/nature06210}

\bibitem[{{Wang} {et~al.}(1997){Wang}, {Qu}, \& {Chen}}]{Wang1997a}
{Wang}, Z.~R., {Qu}, Q.-Y., \& {Chen}, Y. 1997, \aap, 318, L59

\bibitem[{{Wilms} {et~al.}(2000){Wilms}, {Allen}, \& {McCray}}]{Wilms2000a}
{Wilms}, J., {Allen}, A., \& {McCray}, R. 2000, \apj, 542, 914,
  \dodoi{10.1086/317016}

\bibitem[{{Zirakashvili} \& {Aharonian}(2010)}]{Zirakashvili2010a}
{Zirakashvili}, V.~N., \& {Aharonian}, F.~A. 2010, \apj, 708, 965,
  \dodoi{10.1088/0004-637X/708/2/965}

\end{thebibliography}



\end{document}